\documentclass[fleqn,usenatbib]{mnras}

\usepackage{newtxtext,newtxmath}

\usepackage[T1]{fontenc}

\usepackage{graphicx}	
\usepackage{amsmath}	
\usepackage{amssymb}	
\usepackage[para,online,flushleft]{threeparttable}

\font\sevenrm=cmr7
\newcommand{\hii}{H~II}
\newcommand{\hiirs}{\hii~regions}
\newcommand{\uchii}{UC~\hii}
\newcommand{\uchiir}{UC~\hii~region}
\newcommand{\uchiirs}{UC~\hii~regions}
\newcommand{\uchiie}{UC~\hii+EE}
\newcommand{\uchiies}{UC~\hii+EE~regions}

\title[Complete view of UC~H\,{\normalsize \textit{II}} regions with EE]{\textit{Ultracompact~H\,{\normalsize \textit{II}} Regions with Extended Emission: The Complete View}}

\author[E. de la Fuente et al.]{
Eduardo de la Fuente,$^{1}$\thanks{Research Associated at ''HAWC'' Gamma--Ray National Laboratory (CONACyT). Corresponding e--mail: edfuente@gmail.com  }\thanks{Research stay at National Astronomical Observatory of Japan (NAOJ), Tokyo, Japan, (2018).}
Alicia Porras,$^{2}$
Miguel A. Trinidad,$^{3}$
Stanley E. Kurtz,$^{4}$
\newauthor
Simon N. Kemp,$^{1}$ 
Daniel Tafoya,$^{5}$ 
Jos\'e Franco,$^{6}$ 
and Carlos Rodr\'{i}guez--Rico$^{3}$
\\
\\
$^{1}$Instituto de Astronom\'{i}a y Meteorolog\'{i}a, Dpto. de F\'{i}sica, CUCEI, Universidad de Guadalajara, Av. Vallarta 2602, 44130, Guadalajara, Jalisco, M\'exico.
\\
$^{2}$Instituto Nacional de Astrof\'{i}sica, \'Optica y Electr\'onica, Luis E. Erro N\'umero 1, 72840,  Tonantzintla, San Andres Cholula, Puebla, M\'exico.
\\
$^{3}$Departamento de Astronom\'{i}a, Universidad de Guanajuato, Apartado Postal 144, 36000, Guanajuato, Guanajuato, M\'exico. 
\\
$^{4}$Instituto de Radioastronom\'{i}a y Astrof\'{i}sica, UNAM, Antigua Carretera a P\'atzcuaro Numero 8701,  58089, Morelia, Michoac\'an, M\'exico.
\\
$^{5}$Department of Space, Earth and Environment, Chalmers University of Technology, Onsala Space Obsevatory, 439~92 Onsala, Sweden.
\\
$^{6}$Instituto de Astronom\'{i}a, UNAM, Apartado Postal 70--264, CDMX, 04510, M\'exico
}

\date{Accepted XXX. Received YYY; in original for ZZZ}

\pubyear{2019}

\begin{document}
\label{firstpage}
\pagerange{\pageref{firstpage}--\pageref{lastpage}}
\maketitle


\begin{abstract}
In this paper we present the results of a morphological study performed on a sample of 28 ultracompact \hiirs~located near extended free--free emission, using radio continuum observations at 3.6~cm with the C and D VLA~configurations, with the aim of determining a direct connection between them. By using previously published observations in B and D VLA~configurations, we compiled a final catalogue of 21 ultracompact \hiirs ~directly connected with the surrounding extended emission. The observed morphology of most of the ultracompact \hiirs~in radio continuum emission is irregular (single or multi--peaked sources) and resembles a classical bubble structure in the Galactic plane with well--defined cometary arcs. Radio continuum images superimposed on colour composite \textit{Spitzer} images reinforce the assignations of direct connection by the spatial coincidence between the ultracompact components and regions of saturated 24~\micron~emission. We also find that the presence of extended emission may be crucial to understand the observed infrared--excess because an underestimation of ionizing Lyman photons was considered in previous works.
\end{abstract}






\section{Introduction}
\label{intro}

The ultracompact \hii~(\uchii) regions \citep[term coined by][]{IHJ73} are small (size $\leq$ 0.1 pc), dense ($n_{\rm e}$~$\gtrsim$~10$^4$~cm$^{-3}$) regions with high emission measure (EM~$\geq$~10$^7${${\rm pc\ cm}^{-6}$}) and composed of photoionized hydrogen surrounding a recently formed ionizing OB type star \citep[e.g.][]{WC89, K94, Ho07}. They were first studied via interferometric observations \citep[e.g.][]{DH67, Me67, RD67}, and they are thought to represent an intermediate evolutionary state between hypercompact \hiirs~\citep[an earlier phase, e.g.][]{Se06} and compact \hiirs~\citep[a later phase,][]{Ch02}. The physical characteristics of the \uchiirs~have been determined observationally in surveys with the Very Large Array (VLA) performed by numerous authors \citep[see review by][]{Ch02}.

Large-scale (arc--minute) structures of ionized gas have been found that seem to be related to the \uchiirs~\citep{Me67, Co98}. Although these structures were discovered in the 1960s, only a few studies aiming to understand their origin have been conducted{\bf :} \citet{K99}, using the VLA in configuration D at 3.6~cm; \citet{KK01}, using the VLA in configuration DnC at 21~cm; \citet{El05}, using ATCA in configuration 750D at 3.5~cm and \citet{F07}, using \textit{Spitzer} IRAC data \citep[see][for a summary]{F09a, F09b}. Since all these works were complementary and addressed the problem from different points of view, none of them reached conclusive results to clarify the nature of the \uchii~regions. 

\citet{K99} carried out an inspection of 3.6~cm VLA images taken in configurations B \citep{K94} and D, with maximum recoverable scales (MRS) of up to 10--20\arcsec and 3\arcmin, respectively, and  21~cm images from the NVSS survey \citep{Co98} with MRS of up to 10--15\arcmin. This was the first attempt to establish a \textit{direct connection} between the \uchii~regions and the large-scale extended emission (EE), which appears as an arc--minute scale common structure of ionized gas embracing the \uchiir. By comparing the emission distributed over spatial scales of several arc--minutes, they found a common structure in 8 \uchiirs~from a randomly selected sample of 15 sources ($\sim$53\%) from the \citet{K94} survey. 
The confirmation of a direct connection between the UC emission and the EE may impact on \uchiir~ definition, modelling, lifetime problem, and energetics. As \citet{K99} indicate, if this connection is demonstrated, the EE would require about 10 to 20 times more Lyman photons to keep its ionization than the UC emission would require.

 For example, \citet{WC89} and \citet{K94} noted that the IRAS fluxes correspond to a stellar content producing more Lyman continuum photons than is suggested by the radio flux. They explained this disparity by assuming that some fraction of the ionizing photons is absorbed by dust, and/or by the possible presence of a cluster of later-type stars, contributing to the IR luminosity but not to the Lyman continuum flux. If \citet{WC89} and \citet{K94} underestimated the Lyman photon flux as of result of their insensitivity to large-scale structures (i.e., EE) then the need for dust/clusters to explain the IR/radio disparity may have been substantially overestimated.

The present work aims to compile a larger catalogue of \uchiirs~$+$~EE~ (\uchiie) and to determine the significance of the EE for our understanding of the energetics of these regions. Thus, in this paper we present a radio continuum (RC) morphological study using VLA observations of a sample of \uchiirs~with extended emission in order to: i) compare new 3.6~cm VLA images taken in the configuration D with a sample of sources from \citet{KK01} to find (or confirm) the presence of \uchiies; ii) compare  new 3.6~cm VLA images taken in the configuration C with previously published images taken in configuration B and D to probe if the EE has a direct connection with the \uchiir. To do this, we use a multi--resolution image combination method, which we propose to be a technique that works better than the one used by \citet{K99} to search for common and continuous structures at several scales; iii) explain how the presence of EE can impact the IR--excess problem; and iv) enlarge the catalogue of \uchiies. 

In \S~\ref{obs}, we describe the VLA observations, calibration and data reduction process. The VLA low resolution morphological study and \textit{Spitzer} counterparts imagery discussion are presented in \S~\ref{morph}, while a short individual source description is given in \S~\ref{indSources}. The discussion of how the presence of EE can resolve the problem of IR excess is presented in \S~\ref{energy}. Conclusions are given in \S~\ref{summary}.

\section{Observations and Data Reduction}
\label{obs}

\subsection{The sample}
\label{subsample}

The sources targeted in our VLA observations were selected according to the following criteria: i) following \citet{WC89} and \citet{K94}, the sources must exhibit \uchiir~ colours with IR-excess in their IRAS fluxes, and ii) the sources must have EE associated with the \uchiir~in either \citet[][3.6~cm VLA configuration D]{K99} or \citet[][21~cm VLA configuration DnC]{KK01} observations. The resulting sample consists of 28 \uchiies~and they are listed in Table~\ref{tab:tab1}. Additionally, most of the targeted sources have \textit{Spitzer}--IRAC \citep{Faz04} and MIPS \citep{Rie04} 24~\micron~data from GLIMPSE and MIPSGAL surveys \citep{Ch09}. It is worth noting that two of the \uchiirs~seemed not to show a direct connection with the observed EE, thus we propose two other radio continuum peaks to be those the associated with the \uchiirs: i.e. G23.44$-$0.21 instead of G23.46$-$0.20, and G25.71$+$0.04 instead of G25.72$+$0.05. The proposed new \uchiies~are listed in Table~\ref{tab:tab1} and they are discussed in more detail in \S~\ref{discarded}.

\begin{table*}
\caption{The sample of 28 ultracompact \hii ~regions with Extended Emission}
\label{tab:tab1}
\begin{threeparttable}
\begin{tabular}{llccc}
\hline

\uchii & IRAS & R.A.\tnote{a} & Dec. & Distance\tnote{b} \\
region & Source & (J2000) & (J2000) & (kpc) \\

\hline

\, G05.48$-$0.24  & 17559$-$2420  & 17$^{\rm h}$59$^{\rm m}$02\fs9 & $-$24\degr20\arcmin54\farcs{}5 & 14.3$^{(1)}$  \\
\, G05.97$-$1.17  & 18006$-$2422  & 18 03 40.51 & $-$24 22 44.4 & 2.7$^{(2)}$                       \\
\, G10.30$-$0.15  & 18060$-$2005  & 18 08 56.11 & $-$20 05 53.8 & 6.0$^{(3)}$                       \\
\, G12.21$-$0.10  & 18097$-$1825A & 18 12 39.72 & $-$18 24 20.5 & 13.5$^{(2)}$                      \\
\, G18.15$-$0.28  & 18222$-$1321  & 18 25 01.08 & $-$13 15 39.3 & 4.2$^{(3)}$                       \\
\, G19.60$-$0.23  & 18248$-$1158  & 18 27 38.11 & $-$11 56 39.5 & 3.5$^{(2)}$                       \\
\, G23.71$+$0.17  & 18311$-$0809  & 18 33 53.54 & $-$08 07 13.8 & 8.9$^{(2)}$                       \\
\, G23.46$-$0.20\tnote{c}  & 18319$-$0834  & 18 34 44.92 & $-$08 31 07.4 & 9.0$^{(1)}$        \\
\, G25.72$+$0.05\tnote{c}  & 18353$-$0628  & 18 38 02.80 & $-$06 23 47.4 & 9.3$^{(2)}$        \\
\, G28.20$-$0.05  & 18402$-$0417  & 18 42 58.08 & $-$04 14 04.6 & 9.1$^{(3)}$                       \\
\, G31.39$-$0.25  & 18469$-$0132  & 18 49 33.04 & $-$01 29 03.7 & 8.9$^{(5)}$                       \\
\, G33.13$-$0.09  & 18496$+$0004  & 18 52 07.96 & $+$00 08 11.6 & 7.1$^{(6)}$                       \\
\, G35.58$-$0.03  & 18538$+$0216  & 18 56 22.52 & $+$02 20 27.0 & 3.6$^{(6)}$                       \\  
\, G37.55$-$0.11  & 18577$+$0358  & 19 00 16.02 & $+$04 03 15.1 & 9.9$^{(2)}$                       \\
\, G35.20$-$1.74  & 18592$+$0108  & 19 01 46.49 & $+$01 13 24.7 & 3.3$^{(6)}$                       \\
\, G37.87$+$0.40  & 18593$+$0408  & 19 01 53.62 & $+$04 12 49.0 & 9.3$^{(6)}$                       \\
\, G43.24$-$0.04  & 19081$+$0903  & 19 10 33.52 & $+$09 08 25.1 & 11.7$^{(4)}$                      \\
\, G45.07$+$0.13  & 19110$+$1045  & 19 13 22.08 & $+$10 50 53.4 & 6.0$^{(6)}$                       \\
\, G45.12$+$0.13  & 19111$+$1048  & 19 13 27.85 & $+$10 53 36.7 & 6.0$^{(6)}$                       \\
\, G45.45$+$0.06  & 19120$+$1103  & 19 14 21.40 & $+$11 09 14.1 & 6.0$^{(6)}$                       \\
\, G48.61$+$0.02  & 19181$+$1349  & 19 20 30.95 & $+$13 55 26.7 & 9.8$^{(6)}$                       \\
\, G54.10$-$0.06  & 19294$+$1836  & 19 31 42.22 & $+$18 42 51.6 & 7.9$^{(1)}$                       \\
\, G60.88$-$0.13  & 19442$+$2427  & 19 46 20.13 & $+$24 35 29.4 & 2.2$^{(6)}$                       \\
\, G78.44$+$2.66  & 20178$+$4046  & 20 19 39.22 & $+$40 56 36.6 & 3.3$^{(3)}$                       \\
\, G77.96$-$0.01  & 20277$+$3851  & 20 29 36.72 & $+$39 01 21.9 & 4.4$^{(5)}$                       \\
G106.80$+$5.31 & 22176$+$6303  & 22 19 18.23 & $+$63 18 47.5 & 0.9$^{(5)}$                          \\
G111.61$+$0.37 & 23133$+$6050  & 23 15 31.16 & $+$61 07 12.9 & 5.2$^{(3)}$                          \\ 
G111.28$-$0.66 & 23138$+$5945  & 23 16 03.94 & $+$60 02 00.9 & 2.5$^{(3)}$                          \\ 

\hline

\end{tabular}

\begin{tablenotes}
\item[a] Right Ascension (R.A.) and Declination (Dec.) were precessed from 1950 coordinates of \citet{WC89} 
and \citet{K94} papers. Sources are ordered by increasing R.A.\\
\item[b] Distance values are from: (1) \citet{WC89}; (2) \citet{CWC90}; (3) \citet{K94}; (4) \citet{Wa97}; 
(5) \citet{K99}; (6) \citet{Ar02}.\\
\item[c] These sources were previously catalogued as \uchiirs~ \citep{KK01}, but other peaks in our radio 
continuum images (Fig.~\ref{fig:fig1a}) seem to be associated with the EE: G23.44$-$0.21 instead of 
G23.46-0.20, and G25.69$+$0.03 instead of G25.72$+$0.05. See \S~\ref{re_assign} for details.

\end{tablenotes}
\end{threeparttable}
\end{table*}

\subsection{Observations}

Snapshot-mode interferometric observations were carried out with the VLA of the National Radio Astronomy Observatory (NRAO) in the C and D configurations. Continuum emission at 3.6~cm was observed in the C configuration (VLA--C) on December 1st and 13th, 1998, and in the D configuration (VLA--D) on April 3rd, 2005. The total on-source integration times were $\sim$9.5 minutes for the VLA--C observations and $\sim$15 minutes for the VLA--D observations. The total spectral bandwidth was $\Delta\nu$~=~50~MHz for all the observations. As mentioned above, a total of 28 regions were targeted, of which 15 were observed with the VLA--D, 12 with the VLA--C, and one source was observed in both configurations. The absolute flux calibrator 3C 286 (with an assumed flux of 5.23 Jy) was used for all the observations and the quasars J2322+509 (1.60~Jy), J2015+371 (2.95~Jy), J1922+155 (1~Jy), J1832$-$105 (1.28~Jy) were used as phase calibrators. The image noise level (rms), parameters of the resulting synthesized beams and the integrated flux densities of the \uchii~ regions associated with EE are given in Table~\ref{tab:tab2}. In addition, for those sources observed with the VLA--C, images in the configurations B and D were produced using archival data from \citet{K94} and \citet{K99}, respectively, which allowed us to perform multi--resolution image combination\footnote{For three sources, data from only one of those two configurations were available in the archive.}.

\subsection{Data Reduction}

The data were calibrated and reduced with the package AIPS of the NRAO following standard procedures. All the images were self--calibrated in phase in order to improve the signal-to-noise ratio. In order to optimize the compromise between angular resolution and sensitivity, we created the images setting the robust parameter ROBUST = 0 \citep[see][]{Br95}, resulting in an angular resolution ranging from $\sim$~2\arcsec~ to 9\arcsec, which allowed us to detect emission distributed over angular scales up to $\sim$~3\arcmin. Each observational run was calibrated independently and then combined using the AIPS task DBCON. Finally, the multi--resolution--CLEAN algorithm \citep{Wa88} was used to obtain the final multi--resolution images (MRI) by combining the configuration C data with the data of configurations B \citep{K94} and D \citep{K99}. The parameters of the synthesized beam and noise levels of the resulting images are listed in Table~\ref{tab:tab3}.

\begin{table*}
\caption{VLA Observational Parameters and \uchii~ Fluxes}
\label{tab:tab2}
\begin{threeparttable}
\begin{tabular}{llcrrrc}
\hline

\uchii & IRAS & VLA\tnote{a} & S$_{\nu}$\tnote{b} & Beam Size & PA & rms \\
region & Source & configuration & Jy & (arcsec) & (degrees) & mJy beam$^{-1}$ \\

\hline

\, G05.48$-$0.24 & 17559$-$2420  & D & 1.18 & 14.03$\times$7.22 & $-$13 & 0.07   \\
\, G05.97$-$1.17 & 18006$-$2422  & D & 8.06 & 14.07$\times$6.90 & $-$12 & 0.47   \\
\, G10.30$-$0.15 & 18060$-$2005  & D & 5.80 & 12.57$\times$6.98 & $-$12 & 0.35   \\
\, G12.21$-$0.10 & 18097$-$1825A & D & 1.30 & 12.50$\times$7.30 & $-$13 & 0.15   \\
\, G18.15$-$0.28 & 18222$-$1321  & D & 4.20 & 11.52$\times$7.80 & $-$14 & 0.10   \\
\, G19.60$-$0.23 & 18248$-$1158  & B & 2.84 & 1.05$\times$0.76  & 9 & 0.40 \\  
\,               &               & D & 4.33 & 11.28$\times$7.89 & 5 & 1.50 \\
\, G23.71$+$0.17 & 18311$-$0809  & D & 1.49 & 10.61$\times$7.72 & $-$17 & 0.07   \\
\, G23.44$-$0.21\tnote{c}  & 18319$-$0834  & D & 1.16 & 10.70$\times$8.22 & $-$15 & 0.09   \\
\, G25.71$+$0.04\tnote{c}  & 18353$-$0628  & D & 0.68 & 10.93$\times$7.80 & $-$11 & 0.09   \\
\, G28.20$-$0.05 & 18402$-$0417  & D & 0.48 & 9.60$\times$7.87  & $-$08 & 0.06   \\
\, G31.39$-$0.25 & 18469$-$0132  & B & 0.04 & 0.98$\times$0.73  & $-$26 & 0.05 \\
\,               &               & C & 0.65 & 3.05$\times$2.36  & $-$12 & 0.08   \\
\,               &               & D & 0.65 & 9.81$\times$7.83  & 13 & 0.15   \\
\, G33.13$-$0.09 & 18496$+$0004  & B & 0.45 & 0.92$\times$0.74  & $-$29 & 0.15   \\  
\,               &               & C & 0.54 & 2.82$\times$2.37  & $-$14 & 0.15   \\
\,               &               & D & 0.55 & 9.44$\times$7.93  & 16 & 1.80   \\
\, G35.20$-$1.74 & 18592$+$0108  & B & 2.57 & 0.87$\times$0.83  & 17    & 0.90   \\
\,               &               & C & 10.96 & 2.85$\times$2.40  & $-$19 & 0.70   \\
\,               &               & D & 11.31 & 9.85$\times$7.80  & $-$2  & 2.00   \\
\, G35.58$-$0.03 & 18538$+$0216  & B & 0.10 & 0.86$\times$0.73  & $-$20 & 0.07   \\
\,               &               & C & 0.69 & 2.76$\times$2.34  & $-$13 & 0.14   \\
\,               &               & D & 0.85 & 9.12$\times$7.89  &  18   & 0.25   \\
\, G37.55$-$0.11 & 18557$+$0358  & D & 0.93 & 8.79$\times$7.79  & $-$12 & 0.07   \\
\, G37.87$+$0.40 & 18593$+$0408  & B & 2.85 & 0.95$\times$0.79  & 34 & 0.47   \\
\,               &               & C & 4.05 & 2.02$\times$1.86  & $-$14 & 0.30   \\
\,               &               & D & 4.11 & 9.00$\times$7.70  & 6 & 0.50   \\
\, G43.24$-$0.04 & 19081$+$0903  & B & 0.18 & 0.78$\times$0.69  & 7 & 0.15   \\
\,               &               & C & 0.70 & 2.61$\times$2.31  & $-$14 & 0.28   \\
\,               &               & D & 0.73 & 8.57$\times$6.43  & 7 & 0.90   \\
\, G45.07$+$0.13 & 19110$+$1045  & D & 0.73 & 8.36$\times$7.67  & $-$18 & 0.30   \\
\, G45.12$+$0.13 & 19111$+$1048  & D & 1.80 & 8.36$\times$7.67  & $-$18 & 0.30   \\
\, G45.45$+$0.06 & 19120$+$1103  & D & 4.50 & 8.23$\times$7.63  & $-$25 & 0.40   \\
\, G48.61$+$0.02 & 19181$+$1349  & B & 0.05 & 0.90$\times$0.76  & 48 & 0.05   \\
\,               &               & C & 0.82 & 2.48$\times$2.49  & $-$10 & 0.20   \\
\,               &               & D & 1.64 & 8.63$\times$7.57  & 24 & 0.70   \\
\, G54.10$-$0.06 & 19294$+$1836  & D & 0.52 & 8.78$\times$7.77  & $-$24 & 0.06   \\
\, G60.88$-$0.13 & 19442$+$2427  & B & 0.09 & 0.74$\times$0.68  & $-$14 & 0.08   \\
\,               &               & C & 0.49 & 2.42$\times$2.26  & 4 & 0.06   \\
\, G77.96$-$0.01 & 20277$+$3851  & B & 0.08 & 1.65$\times$1.65  & $-$45 & 0.07   \\
\,               &               & C & 0.60 & 2.47$\times$2.28  &    22 & 0.07   \\
\,               &               & D & 0.95 & 7.71$\times$7.42  &    21 & 0.10   \\
\, G78.44$+$2.66 & 20178$+$4046  & B & 0.03 & 0.79$\times$0.66  & $-$58 & 0.05   \\
\,               &               & C & 0.07 & 1.92$\times$1.75  &    30 & 0.08   \\
\,               &               & D & 0.07 & 7.77$\times$7.39  & $-$3 & 0.06   \\
G106.80$+$5.31   & 22176$+$6303  & B & 0.03 & 0.86$\times$0.61  & $-$46 & 0.05   \\
                 &               & C & 0.02 & 9.15$\times$7.28  &    53 & 0.05   \\
G111.28$-$0.66   & 23138$+$5945  & B & 0.05 & 0.86$\times$0.62  & $-$46 & 0.05   \\
                 &               & C & 0.25 & 3.41$\times$2.11  &    75 & 0.06   \\
                 &               & D & 0.30 & 9.33$\times$7.34  &    61 & 0.06   \\
G111.61$+$0.37   & 23133$+$6050  & B & 0.27 & 0.92$\times$0.86  &    14 & 0.65   \\
                 &               & C & 0.93 & 3.32$\times$2.11  &    70 & 0.07   \\
                 &               & D & 0.97 & 9.29$\times$7.36  &    59 & 0.20   \\

\hline

\end{tabular}

\begin{tablenotes}
\item[a] For those sources observed with the VLA configuration C, VLA configurations B and D images 
were re-created using data from \citet{K94} and \citet{K99}, respectively.\\
\item[b] \uchii~ integrated flux densities were obtained using suitable boxes covering the whole 
radio continuum emission above 3 times the noise level.\\
\item[c] These sources are proposed to be related to the EE. See \S~\ref{re_assign} for details.

\end{tablenotes}
\end{threeparttable}
\end{table*}

\begin{table*}
\caption{Combined multi--configuration VLA image parameters}
\label{tab:tab3}
\begin{threeparttable}
\begin{tabular}{cccccc}
\hline

\uchii & IRAS\tnote{a} & VLA & Beam Size & PA & rms \\
region & Source & configurations\tnote{c}& (arcsec) & (degrees) & mJy beam$^{\rm -1}$ \\

\hline
\,\,\,\, G19.60$-$0.23\tnote{b} & 18248$-$1158  & B\phantom{C}D  & 1.05$\times$0.76 & \phantom{$-$0}9 & 0.360  \\
\, G31.39$-$0.25 & 18469$-$0132  & BCD & 0.98$\times$0.73 & $-$26 & 0.045  \\
\, G33.13$-$0.09 & 18496$+$0004  & BCD & 0.92$\times$0.74 & $-$29 & 0.065  \\
\, G35.58$-$0.03 & 18538$+$0216  & BCD & 0.86$\times$0.73 & $-$20 & 0.070  \\
\, G35.20$-$1.74 & 18592$+$0108  & BCD & 0.87$\times$0.83 & \phantom{$-$}17 & 0.300  \\
\, G37.87$+$0.40 & 18593$+$0408  & BCD & 0.95$\times$0.79 & \phantom{$-$}34 & 0.300  \\
\, G43.24$-$0.04 & 19081$+$0903  & BCD & 0.78$\times$0.69 & \phantom{$-$0}7 & 0.400  \\
\, G48.61$+$0.02 & 19181$+$1349  & BCD & 0.90$\times$0.76 & \phantom{$-$}48 & 0.270  \\
\, G60.88$-$0.13 & 19442$+$2427  & BC\phantom{D}  & 0.74$\times$0.68 & $-$14 & 0.060  \\
\, G78.44$+$2.66 & 20178$+$4046  & BCD & 0.79$\times$0.66 & $-$58 & 0.055  \\
\, G77.96$-$0.01 & 20277$+$3851  & BCD & 1.65$\times$1.65 & $-$45 & 0.100  \\
  G106.80$+$5.31 & 22176$+$6303  & BC\phantom{D}  & 0.86$\times$0.61 & $-$46 & 0.050  \\    
  G111.61$+$0.37 & 23133$+$6050  & BCD & 0.92$\times$0.86 & \phantom{$-$}14 & 0.040  \\
  G111.28$-$0.66 & 23138$+$5945  & BCD & 0.86$\times$0.62 & $-$46 & 0.060  \\
\hline

\end{tabular}

\begin{tablenotes}
\item[a] \uchiirs ~are ordered by IRAS name.

\item[b] The multi--resolution image of G19.60$-$0.23 is not shown in Fig.~\ref{fig:fig2a} but is shown superimposed on the \textit{Spitzer} image in Fig.~\ref{fig:fig3a}.

\item[c] VLA configurations combined to produce the multi--resolution images.

\end{tablenotes}
\end{threeparttable}
\end{table*}

\section{Morphological Study}
\label{morph}

\subsection{Relation between \uchiir~and the extended emission}
\label{dircon}

The radio continuum images of the 15 regions taken in the VLA--D at 3.6~cm are shown in  Fig.~\ref{fig:fig1a}. A comparison between the VLA--D 3.6~cm images with those at 21~cm from \citet{KK01} reveals that the emission has a similar morphology, albeit at a different angular scale (MRS of 3\arcmin~and 15\arcmin, respectively). Therefore, as discussed in \citet{K99} and \citet{KK01}, these sources are considered to be systems composed of an \uchiir~ and EE. 

\begin{figure*}
    \includegraphics[width=0.91\textwidth]{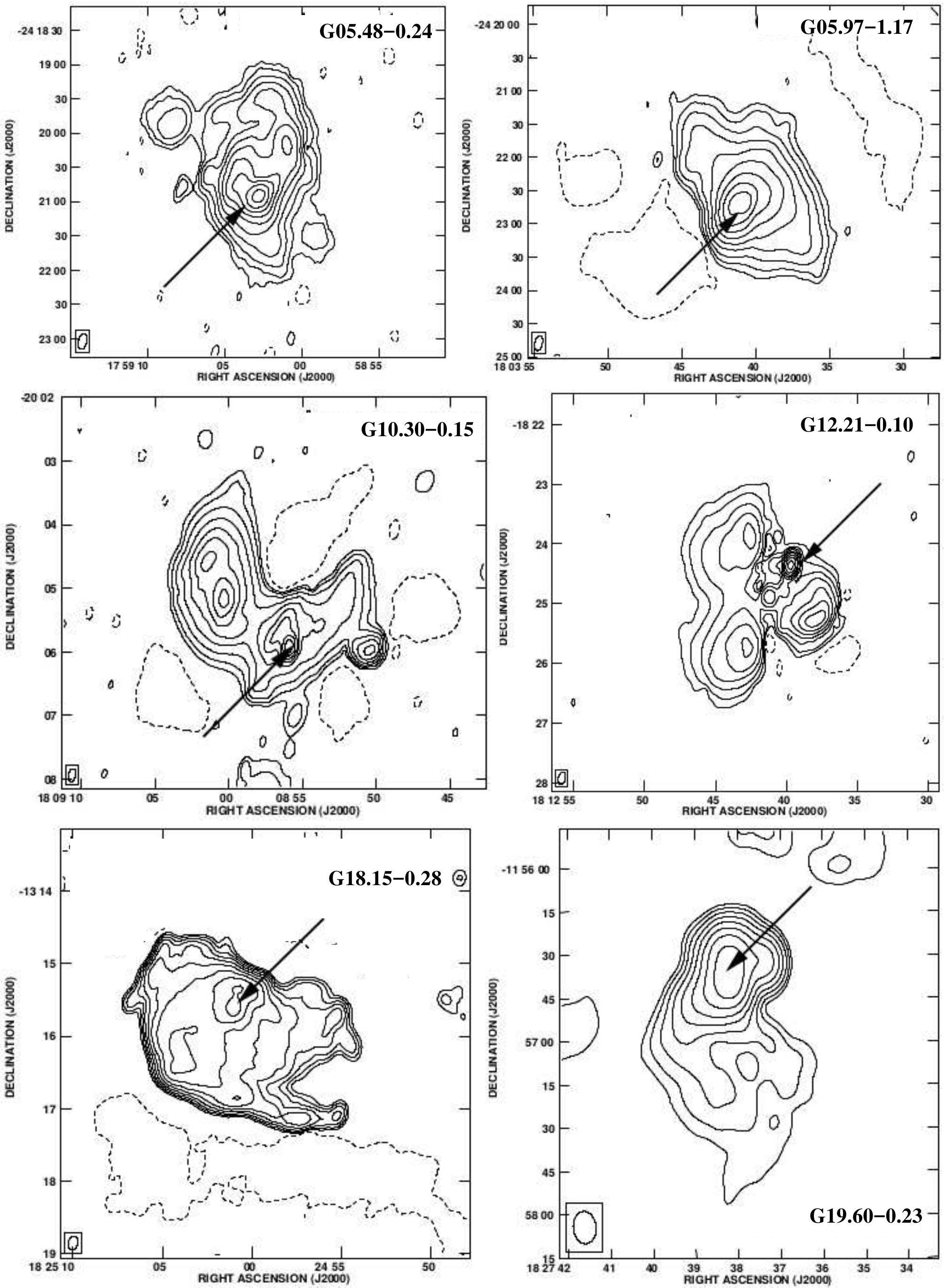}

  \caption{Images at 3.6~cm of the 15 \uchiies~observed in the VLA--D. The contour values are integer 
  multiples of $-$3 and 3 times the image noise level (mJy beam$^{-1}$), which is listed in Table~\ref{tab:tab2}. The solid arrows indicate the position of the \uchiir~associated with the EE from previous works. For two regions (G23.46$-$0.20 and G25.72$+$0.05), a dashed arrow indicates the position of the new \uchiir~ proposed in this work to be associated with the EE. See \S~\ref{re_assign} for details.}
  \label{fig:fig1a}
\end{figure*}

  \setcounter{figure}{0}
\begin{figure*}
  \includegraphics[width=0.91\textwidth]{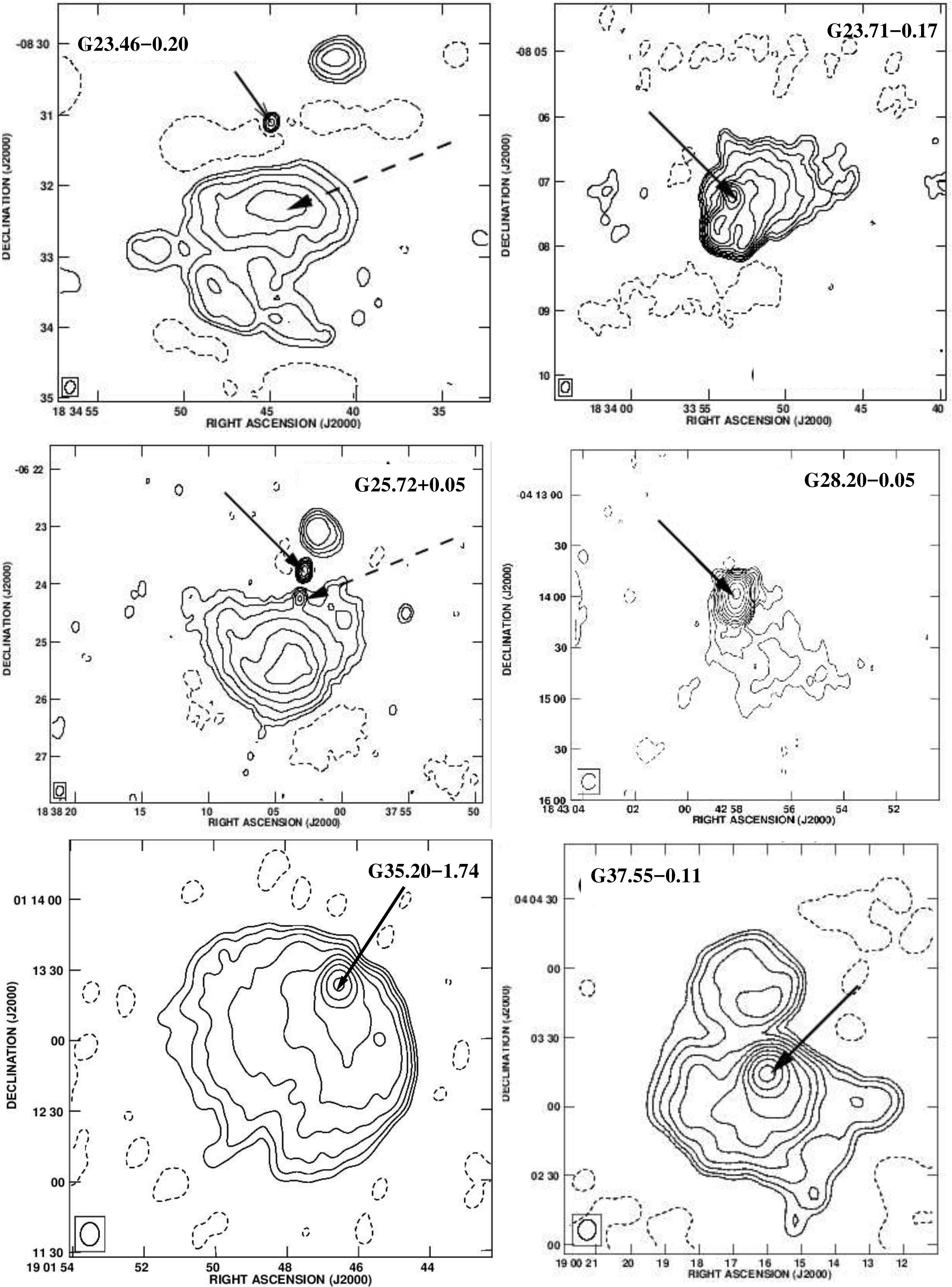}

  \caption{\textit{continued.}}
  \label{fig:fig1b}
\end{figure*}

  \setcounter{figure}{0}
\begin{figure*}
  \includegraphics[width=0.91\textwidth]{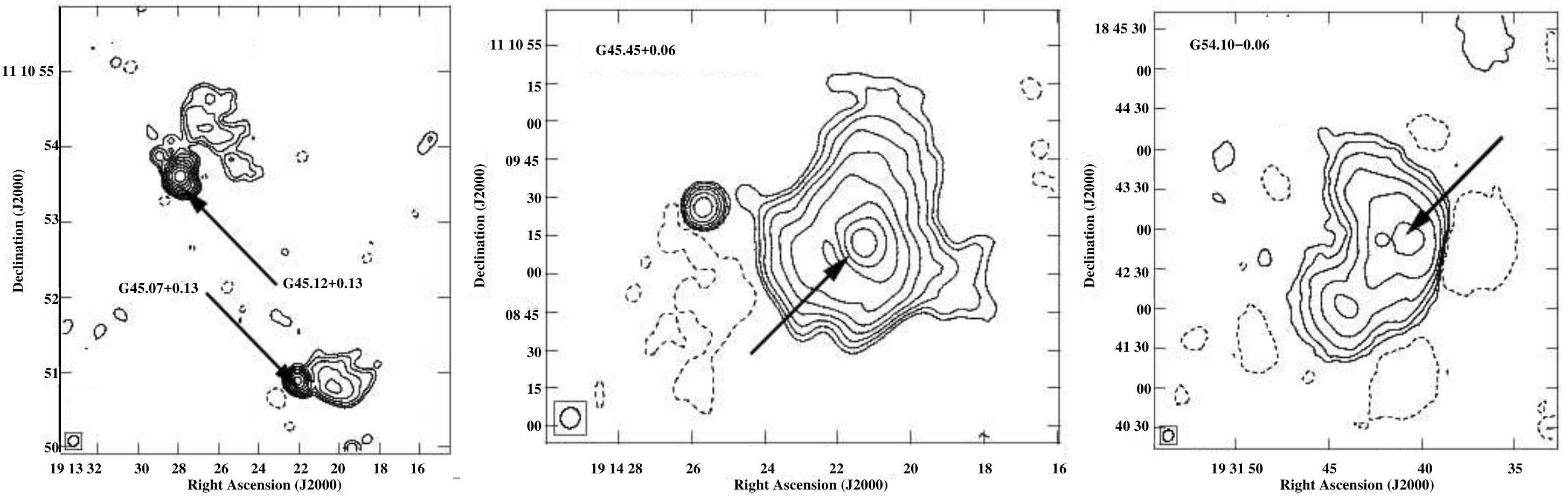}

  \caption{\textit{continued.}}
  \label{fig:fig1c}
\end{figure*}

Fig.~\ref{fig:fig2a} shows images at 3.6~cm taken in the VLA--C and their respective MRIs obtained after combining data from the available configurations (see Table~\ref{tab:tab3} for details). If the multi--configuration image shows that the UC emission is completely embedded within the EE, there is a strong indication that both emissions are directly connected. A good example of such a case is G35.20--1.74. The emission of this source exhibits an extended smoothly distributed morphology in the VLA--D image (Fig.~\ref{fig:fig1a}), as well as in its VLA--C and VLA--(B+C+D) images (Fig.~\ref{fig:fig2a}). This way of determining the direct connection between \uchiir~ and the EE is more robust than the simple inspection made by \citet{K99} because if there is emission filtered out in the observation of one configuration, the MRIs can reveal the association. It is important to remark that \citet{K99} and \citet{KK01} discarded \uchiie~ candidates due to the emission not being continuous, our observations recovered the missing emission for those sources and the connection between \uchiir~ and the EE was revealed. Following this technique, we completely ruled out 9 candidate sources, marked with 'no' or 'unlikely' labels by \citet{KK01} and \citet{K99} (see Table~\ref{tab:tab4}). The criterion to rule out these candidates was that they exhibit a clearly disconnected distribution between the UC emission and the EE. An example of this kind of behaviour is G48.61$+$0.02; although it has continuous extended emission on its VLA--D image \citep[see Figure 7 of][]{K99}, this behaviour is not observed either on  the VLA--C or on VLA--BCD images (Fig.~\ref{fig:fig2a}), confirming the conclusions of \citet{K99}. All discarded sources are presented and commented on \S~\ref{discarded}. Our results on direct--connection between UC and EE emissions are summarized in Table~\ref{tab:tab4}. 

\begin{figure*}
    \includegraphics[width=0.94\textwidth]{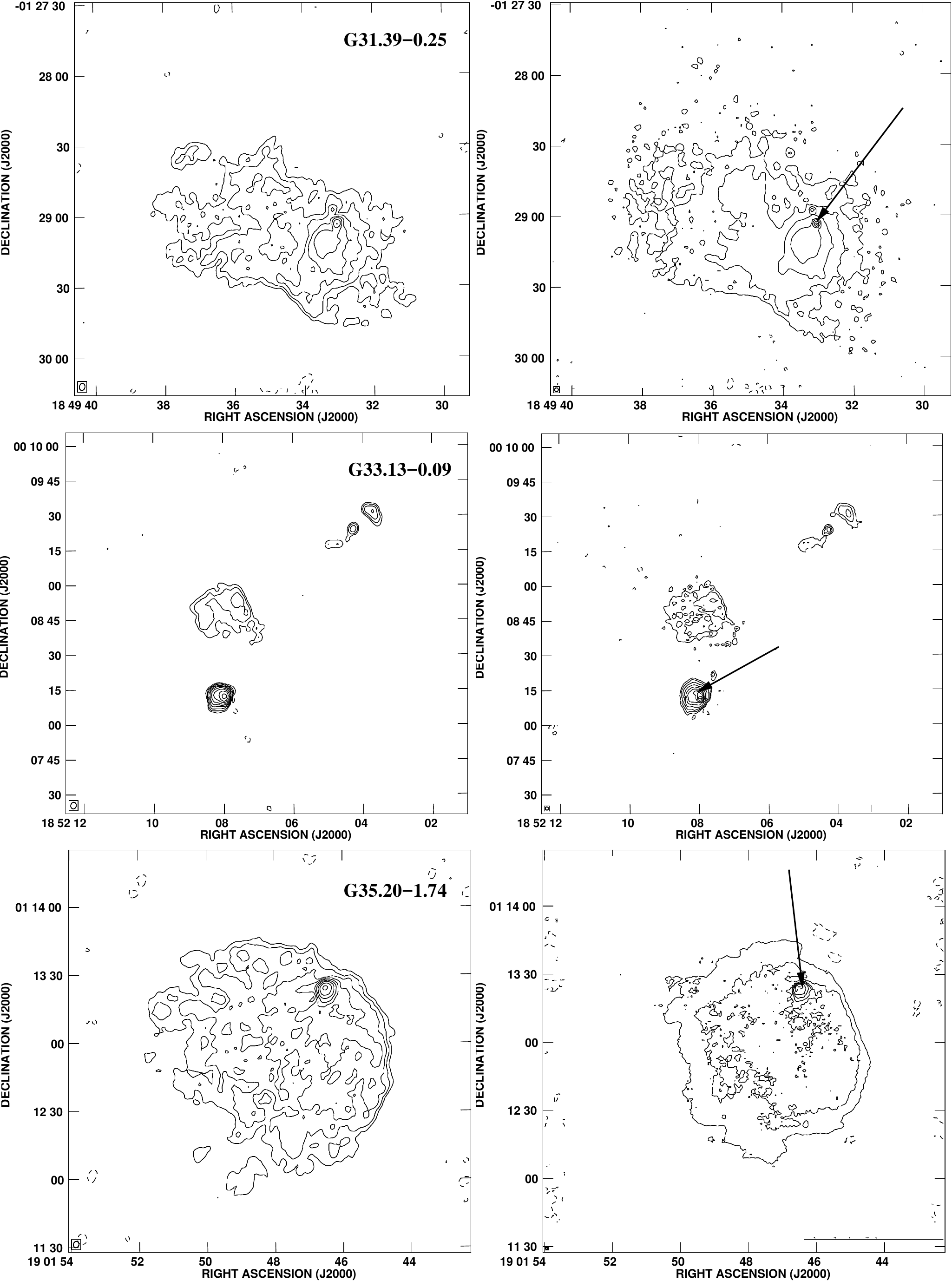}

  \caption{New VLA--C configuration images at 3.6~cm (\textit{left}) and the Multi--Resolution images combining available B, C and D VLA configuration images at 3.6~cm (\textit{right}), as specified in Table~\ref{tab:tab3}. The contour values are integer multiples of $-$3 and 3 times the image noise level (mJy beam$^{-1}$) listed in Table~\ref{tab:tab2} and Table~\ref{tab:tab3}, respectively. The solid arrows show the position of the \uchiirs.}
  \label{fig:fig2a}
\end{figure*}

  \setcounter{figure}{1}
\begin{figure*}
 \includegraphics[width=0.94\textwidth]{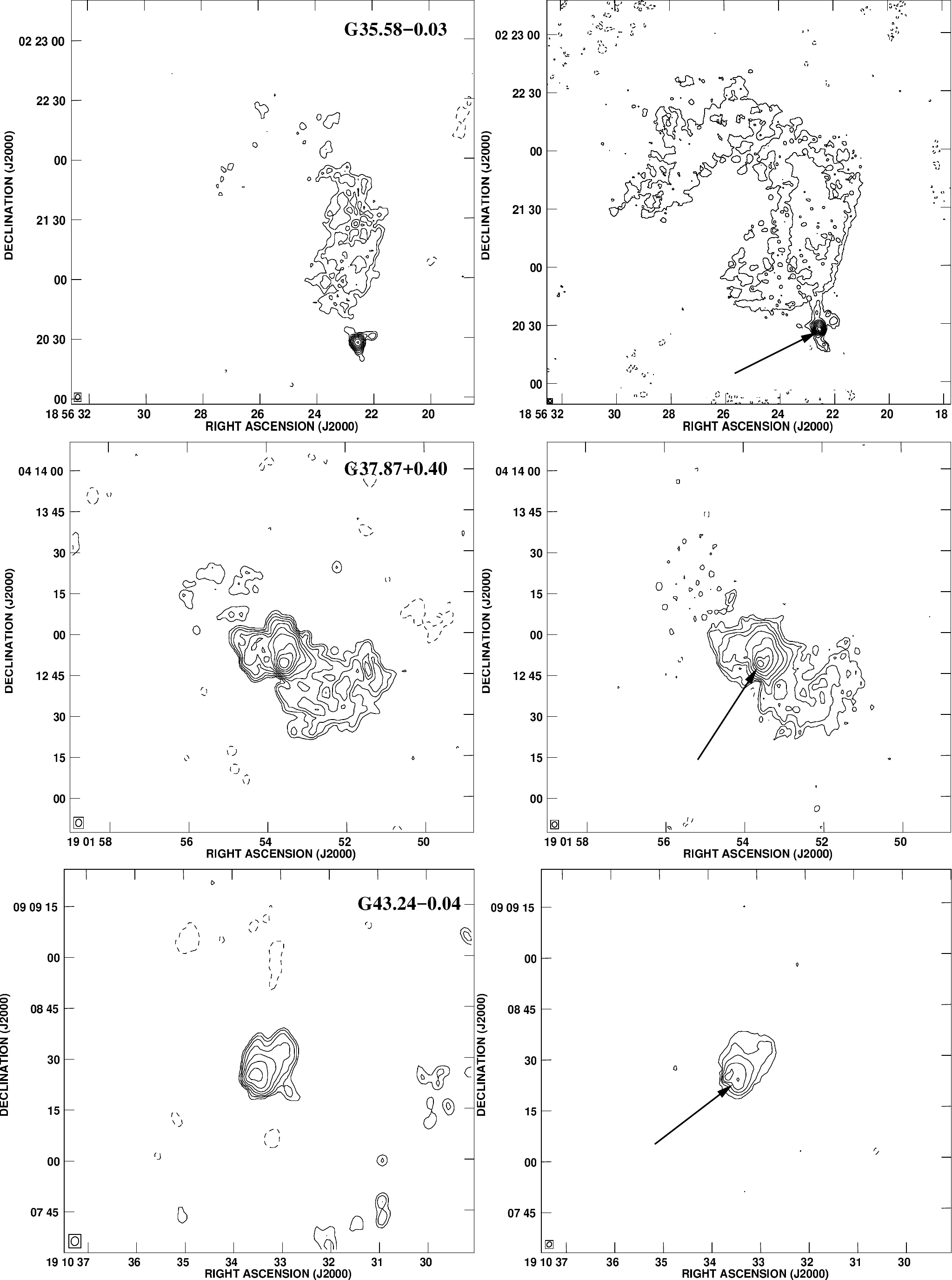}

  \caption{\textit{continued.}}
  \label{fig:fig2b}
\end{figure*}

\

  \setcounter{figure}{1}
\begin{figure*}
  \includegraphics[width=0.94\textwidth]{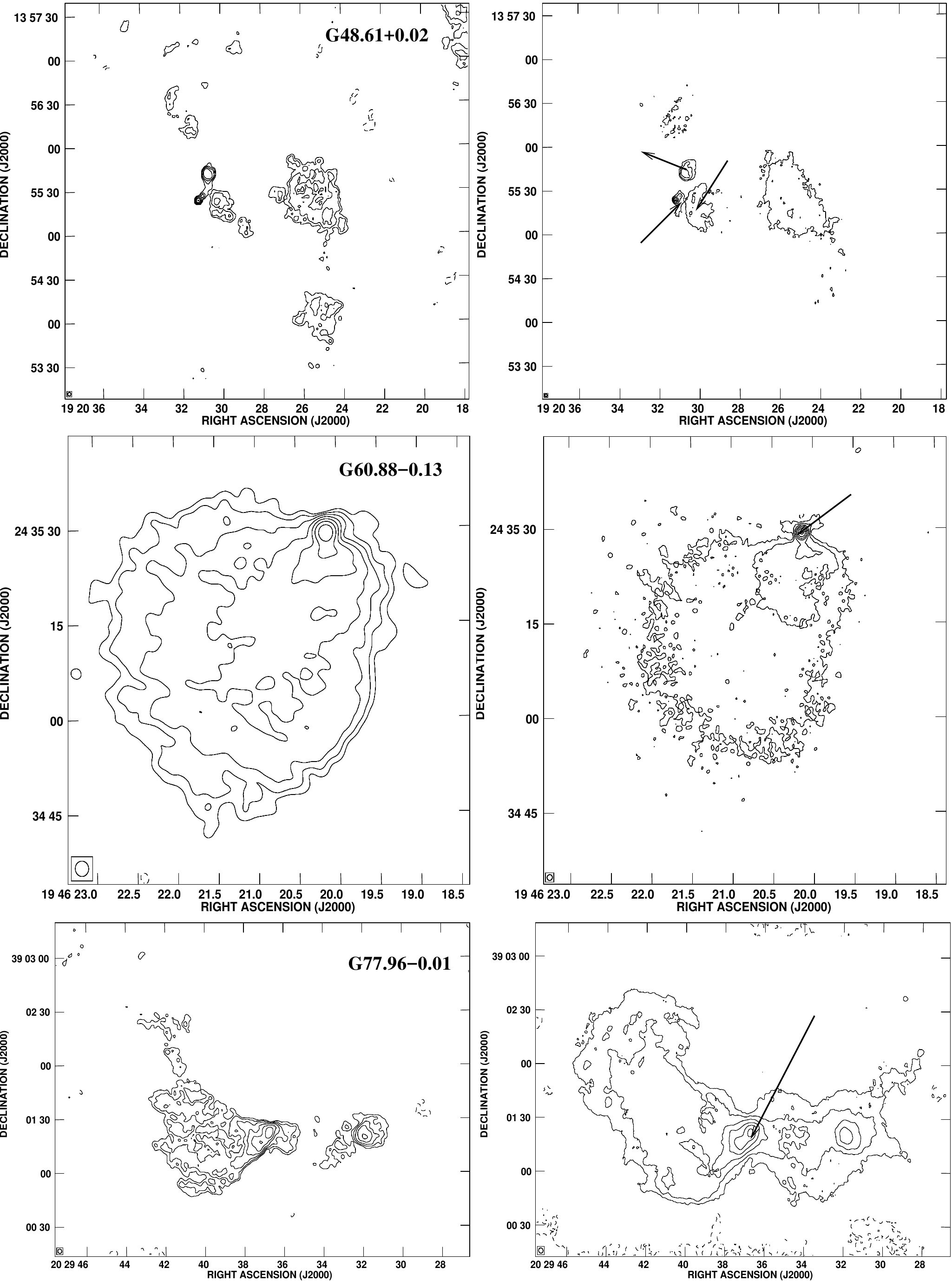}

  \caption{\textit{continued.}}
  \label{fig:fig2c}
\end{figure*}

  \setcounter{figure}{1}
\begin{figure*}
    \includegraphics[width=\textwidth]{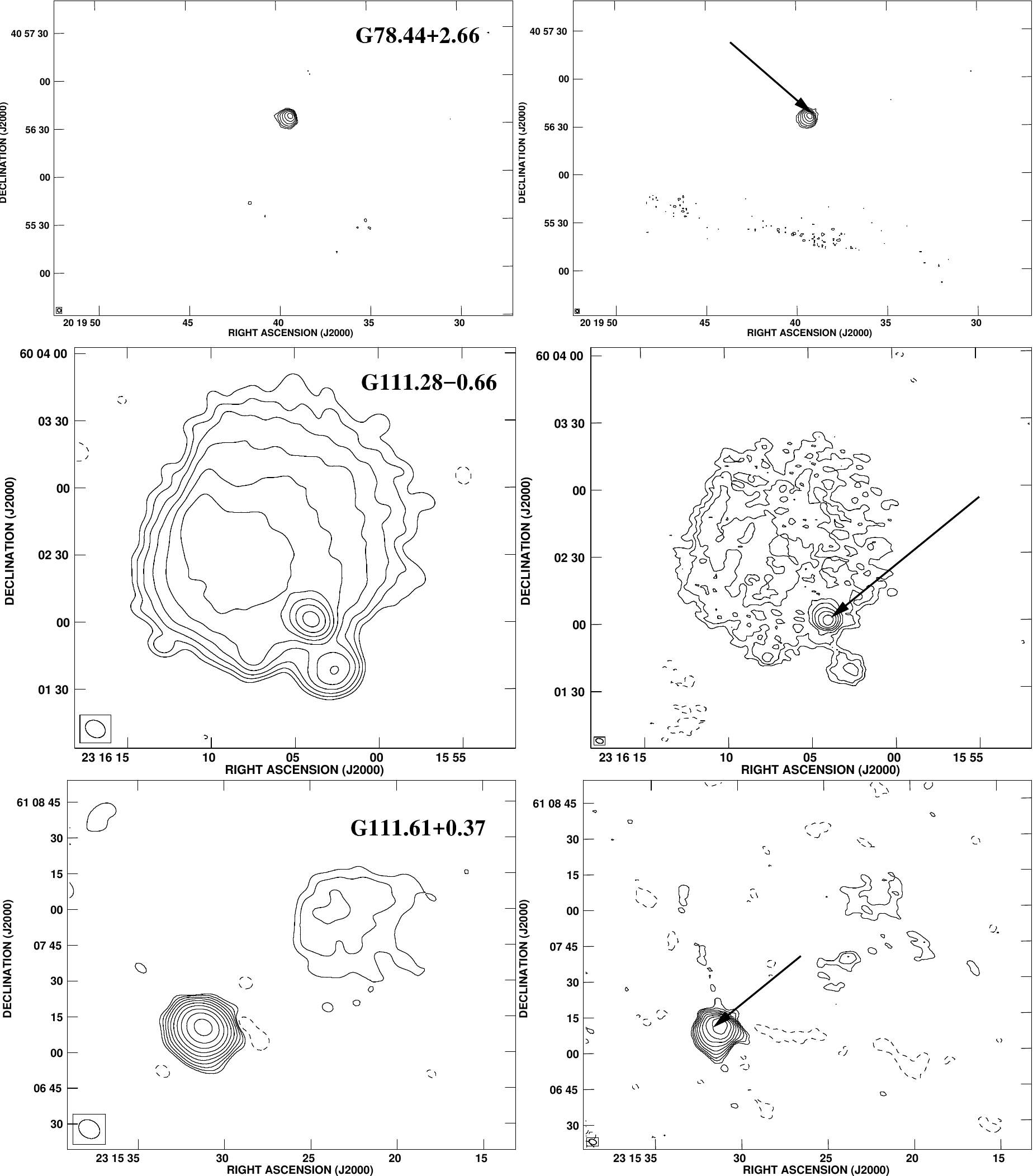}

  \caption{\textit{continued.}}
  \label{fig:fig2d}
\end{figure*}

  \setcounter{figure}{1}
\begin{figure*}
    \includegraphics[width=\textwidth]{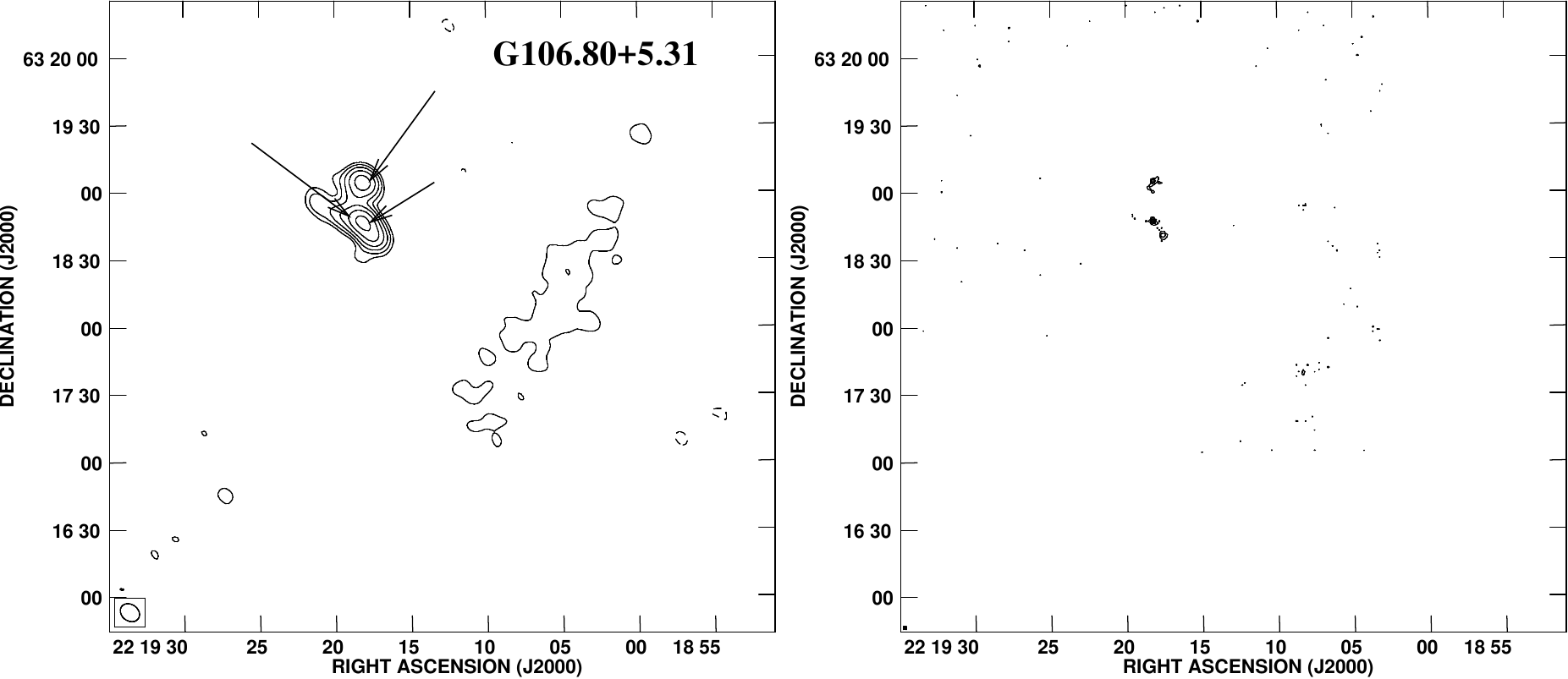}

  \caption{\textit{continued.}}
  \label{fig:fig2e}
\end{figure*}

We perform a morphological classification similar to that proposed by \citet{WC89}, based on shapes from radio--continuum emission. According to them, morphological types are: I=spherical (or unresolved), II=cometary (parabolic), III=core--halo, IV=shell, V=irregular (or multiply--peaked). Based on Figs.~\ref{fig:fig1a} and \ref{fig:fig2a} of VLA emission at 3.6~cm, we associate a morphological type to each source (see Table~\ref{tab:tab4}). We find that cometary type is dominant in our sample (43\%), morphological types spherical and irregular coincide in number (21.5\%), while the least favoured types are core-halo (16\%) and shell (0\%).

The radio continuum morphology presented in this work traces ionized gas and matches the observed IRAC morphology (see \S~\ref{Ch_bubbles}), where dust and Polycyclic aromatic hydrocarbons (PAH$'$s) features on the cometary arcs are clearly observed and could be related to photo--dissociated regions \citep[PDR$'$s,][]{HT97}. It is important to remark that the IRAC morphology at meso--scales, avoiding the spatial filter produced by interferometers at large scale{\bf s}, is in agreement with the radio continuum morphology presented in this work{\bf ;} supporting results are shown in Tables~\ref{tab:tab4} and \ref{tab:tab5}.   

\begin{table*}
\caption{Results from Morphological study of Extended Emission}
\label{tab:tab4}
\begin{threeparttable}
\begin{tabular}{clcccccccc}

\hline

\multicolumn{1}{c}{\uchii~} & \multicolumn{1}{c}{EE\tnote{a}} & \multicolumn{2}{c}{Direct Connection\tnote{a}} & \multicolumn{1}{c}{EE Size\tnote{b}} & \multicolumn{1}{c}{RC} & \multicolumn{1}{c}{Morphological} & \multicolumn{3}{c}{{\it Spitzer} Bubble}  \\
\multicolumn{1}{c}{region} &  & \multicolumn{1}{c}{previous} & \multicolumn{1}{c}{this work} &\multicolumn{1}{c}{approx} & \multicolumn{1}{c}{Peaks} & \multicolumn{1}{c}{type\tnote{c}} & \multicolumn{1}{c}{Ch06/Ch07\tnote{d}} & \multicolumn{1}{c}{Simp12\tnote{e}} & \multicolumn{1}{c}{this work} \\

\hline

\,G05.48$-$0.24     & Y & {\bf Y (1)} & Y & $3.0' \times 2.0'$     & 5 & III & Y & Y & Y \\
\,G05.97$-$1.17     & Y & Y (3) & Y & $3.5' \times 2.0'$           & 1 &  II & N & N & Y \\  
\,G10.30$-$0.15     & Y & Y (3) & Y & $4.0' \times 1.5'$           & 5 &  V  & Y & Y & Y \\ 
\,G12.21$-$0.10     & Y & Y (3) & Y & $4.0' \times 3.0'$           & 5 &  V  & N & Y & Y \\ 
\,G18.15$-$0.28     & Y & ---   & Y & $2.0' \times 4.0'$           & 2 &  II & Y & N & Y \\
\,G19.60$-$0.23     & Y & ---   & Y & $0.8' \times 2.0'$           & 5 &  II & N & Y & Y \\   
\,G23.44$-$0.21     & Y & {\bf ---} & Y & $2.0' \times 2.0'$       & 1 &  II & N & Y & Y \\
\,G23.71$+$0.17     & Y & Y (3) & Y & $2.0' \times 2.0'$           & 1 &  II & Y & Y & Y \\ 
\,G25.69$+$0.03 & Y & {\bf ---} & Y & $3.0' \times 4.0'$     & 2 &  II & Y & N & Y \\ 
\,G28.20$-$0.05     & Y & ---   & N & $1.5' \times 1.0'$           & 1 & III & Y & N & Y \\ 
\,G31.39$-$0.25     & Y & P (2) & Y & $2.5' \times 1.5'$           & 2 &  V  & Y & Y & P \\ 
\,G33.13$-$0.09     & Y & U (2) & N & $0.5' \times 0.5'$           & 1 &  I  & Y & N & N \\
\,G35.20$-$1.74     & Y & ---   & Y & $2.0' \times 2.0'$           & 1 &  II & N & N & N \\
\,G35.58$-$0.03     & Y & U (2) & N & $2.0' \times 2.0'$           & 1 &  I  & Y & N & Y \\
\,G37.55$-$0.11     & Y & Y (3) & Y & $2.0' \times 2.0'$           & 2 & III & N & Y & Y \\
\,G37.87$+$0.40     & Y & P (2) & Y & $2.0' \times 0.5'$           & 1 &  II & Y & N & N \\
\,G43.24$-$0.04     & N & N (2) & N & $0.5' \times 0.5'$           & 1 &  II & N & N & N \\
\,G45.07$+$0.13     & Y & ---   & Y & $1.0' \times 1.0'$           & 1 &  I  & Y & Y & Y \\
\,G45.12$+$0.13     & Y & ---   & P & $1.5' \times 1.5'$           & 1 &  I  & N & Y & Y \\
\,G45.45$+$0.06     & Y & ---   & Y & $2.0' \times 2.0'$           & 1 & III & N & N & Y \\
\,G48.61$+$0.02     & Y & P (2) & N & $2.0' \times 1.5'$           & 3 &  V  & Y & Y & N \\
\,G54.10$-$0.06     & Y & ---   & Y & $3.0' \times 2.0'$           & 2 &  II & Y & Y & Y \\
\,G60.88$-$0.13     & Y & P (2) & Y & $1.0' \times 3.0' $          & 1 &  II & Y & Y & Y \\
\,G77.96$-$0.01     & Y & P (2) & Y & $2.0' \times 3.0'$           & 2 &  V  & N & N & N \\
\,G78.44$+$2.66     & Y & U (2) & N & $0.3'\times 0.3'$            & 1 &  I  & N & N & N \\
G106.80$+$5.31      & Y & U (2) & N & $1.0' \times 1.0'$           & 2 &  V  & N & N & N \\
G111.28$-$0.66      & Y & P (2) & Y & $2.5' \times 2.5'$           & 2 &  II & N & N & N \\
G111.61$+$0.37      & N & N (2) & N & $2.5' \times 2.5'$           & 2 &  I  & N & N & N \\

\hline

\end{tabular}

\begin{tablenotes}
\item[a] Y = yes, N = no, P = possible, U = unlikely, in previous works according to: (1) \citet{Ko96}, (2) \citet{K99} and (3) \citet{KK01}.

\item[b] Presence of extended emission considering continuum images at arc--min scales with VLA--D configuration. EE size was obtained fitting a box to the contour of lower value on 3.6~cm maps.

\item[c] Morphological type according \citet{WC89}: I=spherical (or unresolved), II=cometary (parabolic), III=core--halo, IV=shell, V=irregular (or multiply--peaked).

\item[d] Presence of Bubble--like structures from IRAC images according to \citet{Ch06, Ch07}.

\item[e] Presence of Bubble--like structures from IRAC and MIPS images according to \citet{Simp12}.

\end{tablenotes}
\end{threeparttable}
\end{table*}

\citet{KK01,KK02,KK03} suggest a model to explain the origin of extended emission considering the hierarchical structure of molecular clouds and a champagne flow from one or more ionizing sources (see their Figure 8). The density gradient expansion that generates EE in a natural way in compact regions is described by \citet{Fr90}. However, in most of the sources that we study in this work the IRAC 8~$\mu$m emission seems to be surrounding the MIPS 24~$\mu$m and radio--continuum emission, which can be more easily interpreted as originating in a bubble as described by \citet{Ch06}. Although, it is important to point out that without information of the kinematics of the ionised gas and density field of the ambient medium we can not completely rule out the Kim \& Koo's champagne model, at least not for all the sources under study in the present work.  A champagne--like flow scenario could be tested source--by--source having the respective molecular and kinematical data.


For completeness and comparison to our sample, the sources with IRAC bubble structures from \citet{Ch06} are indicated in Table~\ref{tab:tab4}. Three--quarters of the bubbles in \citet{Ch06} are associated with B4--B9 stars (too cool to produce detectable radio \hiirs ), while the others are produced by young O--B3 stars with detectable radio \hiirs . They suggest that bubbles that overlap known \hiirs ~are produced by stellar winds and radiation pressure from young OB stars in massive star formation regions. 

In summary, based on the morphology of radio continuum emission images at 3.6~cm, for the 28 objects in the original sample (Table~\ref{tab:tab1}), 9 are discarded as \uchiies~ ('N' at fourth column in Table~\ref{tab:tab4}) resulting in 19 sources confirmed plus 2 new sources (G23.44$-$0.21 and G25.69$+$0.03) giving a final catalogue of 21 \uchiies~ listed in Table~\ref{tab:tab5}. This table also includes observed EE--sizes and the number of radio continuum emission peaks. The average EE--size in these sources is 4.9$\pm$3.1~arcmin$^2$. The radio continuum peaks could be associated with independent \uchiirs~ or IR sources, and the EE related to cometary--like shape{\bf s} at large scales can be the bubble--like structures described by \citet{Ch06}, \citet{F09a} and \citet{PH08} with PDR$'$s on the cometary arcs and where dust, ionized gas, and a possible stellar cluster coexist.

\subsection{\textit{Spitzer} IRAC and 24~\texorpdfstring{$\bmath{\umu}$m}~ (MIPS) emissions}
\label{Ch_bubbles}

\citet{Ch06, Ch07} looking at the IRAC 3.6, 4.5, 5.8 and 8.0~\micron ~bands \citep{Ben03, Faz04} from the GLIMPSE Spitzer and Ancillary Data \citep{Ben03, Ch09}, found and catalogued nearly 600 IR `bubbles' around OB stars in the galaxy. Later, using 24~\micron ~diffuse emission sources from MIPSGAL, \citet{Car09} and \citet{Ban10} found similar structures with spatially coincident 21~cm continuum emission. They discovered several hundreds of new Galactic \hiirs, including those associated with \citet{Ch06} bubbles.

IRAC images are dominated by PAH$'$s at 3.3, 6.2 and 7.7~\micron~ \citep{LiD01}, where the strongest emission is observed at 8.0~\micron~\citep{Ch06}. This band traces knots--like sources that could be either star clusters or externally illuminated condensations. The 5.8~\micron ~band is considered a good dust tracer \citep{Ch04} while the 4.5~\micron ~band, free of dust and PAH$'$s emission, traces ionized gas from Br$\alpha$ and shocked emission \citep{Cy08}. The 3.6~\micron ~band also shows the presence of stellar clusters and both ionized and reflection nebulosities \citep{Fla06, Sm06}. Therefore, GLIMPSE images are particularly helpful in identifying bubbles as emission from PAH$'$s ~at 8.0~\micron , and to detect bubble rims \citep{Simp12}. The MIPS 24~\micron ~emission associated to warm dust is commonly observed located within the bubble and with the morphology that closely traces the radio continuum emission \citep{Ch06, Wat08, Wat10, Simp12}. Such bright 24~\micron~and radio continuum emissions are frequently surrounded by shells traced by intense 8.0~\micron~emission in giant Galactic \hiirs~ and extragalactic star--forming regions \citep[e.g.][]{Pov07, Bdo08, RelKen09, Fla11}. Considering the latter, \citet{Simp12} catalogued 5106 bubbles using RGB images (Red = 24~\micron , Green = 8.0~\micron , and Blue = 4.5~\micron) finding red emission surrounded by green rims or spherical--type shapes.
 
\citet{PH08} imaged 58 compact \hiirs~ using only IRAC bands. They show these sources as isolated entities within larger star--forming regions, many of them with similar structures consisting of spheroidal shells with narrow emission rims, and in many cases, attached to filaments and haloes, that are particularly evident within the 5.8~\micron ~and 8.0~\micron ~bands. In this direction, an IRAC--\textit{Spitzer} study of several \uchiies~ (many of them listed in Table~\ref{tab:tab1}), including IRAC photometry, is reported by \citet{F09b}. In this study, it was found that the observed EE looks similar to the `bubbles' detected by \citet{Ch06}, and heated dust \citep[traced by 5.8 and 8.0~\micron ~bands, ][]{Fla06, Sm06, Ku07} coexists with ionized gas \citep[traced by 4.5~\micron ~band, ][]{Cy08} and star clusters \citep[traced by 3.6~\micron ~band, ][]{Fla06, Sm06}.

The morphology of the studied sources in all IRAC bands is similar but increasing in intensity from 3.6 to 8.0~\micron . Also, a visual inspection of these sources in the 24~\micron ~band shows saturation at the position of all the \uchii~ peaks, except for G12.21--0.10, suggesting that this MIPS band is useful to locate candidate \uchiies. In these \uchiies, the 24~\micron ~emission is more intense than in the brightest IRAC band (8.0~\micron ) indicating the presence of relatively cool dust. 

\textit{Spitzer} RGB images (Red = 24~\micron , Green = 8.0~\micron , and Blue = 4.5~\micron ) of all \uchiies~ in our catalogue are shown in Fig.~\ref{fig:fig3a}. These images show a similar behaviour to that noted by \citet[][and references therein]{Simp12}: the MIPS emission clearly coincides with the VLA radio continuum emission and green (8.0~\micron)  rims surround them. The 8.0~\micron ~emission traces very well the cometary arcs observed in radio continuum. The cometary arcs could be PDR$'$s with PAH$'$s \citep[][and references therein]{HT97, DMFSh98, Tie04, Peet05}, but HI observations are needed to confirm their PDR nature. It is clear that \textit{Spitzer} images are good to find structures at meso--scales with better resolution than radio--continuum, and the saturated position of the \uchiirs~ is in agreement with the idea that \uchiirs~ are the brightest objects in the Galaxy at MIR/FIR wavelengths \citep{Ch02}. For example, in the G25.7$+$0.0 complex, the position of the saturated 24~\micron ~emission confirms our suggestion that the observed EE initially associated with G25.72$+$0.05 is instead related to G25.69$+$0.03, a possible compact \hii ~region, and G25.71$+$0.04 which seems to be an \uchiir ~embedded at the edge of this EE (see Fig.~\ref{fig:fig3a}).

\begin{figure*}
    \includegraphics[width=\textwidth]{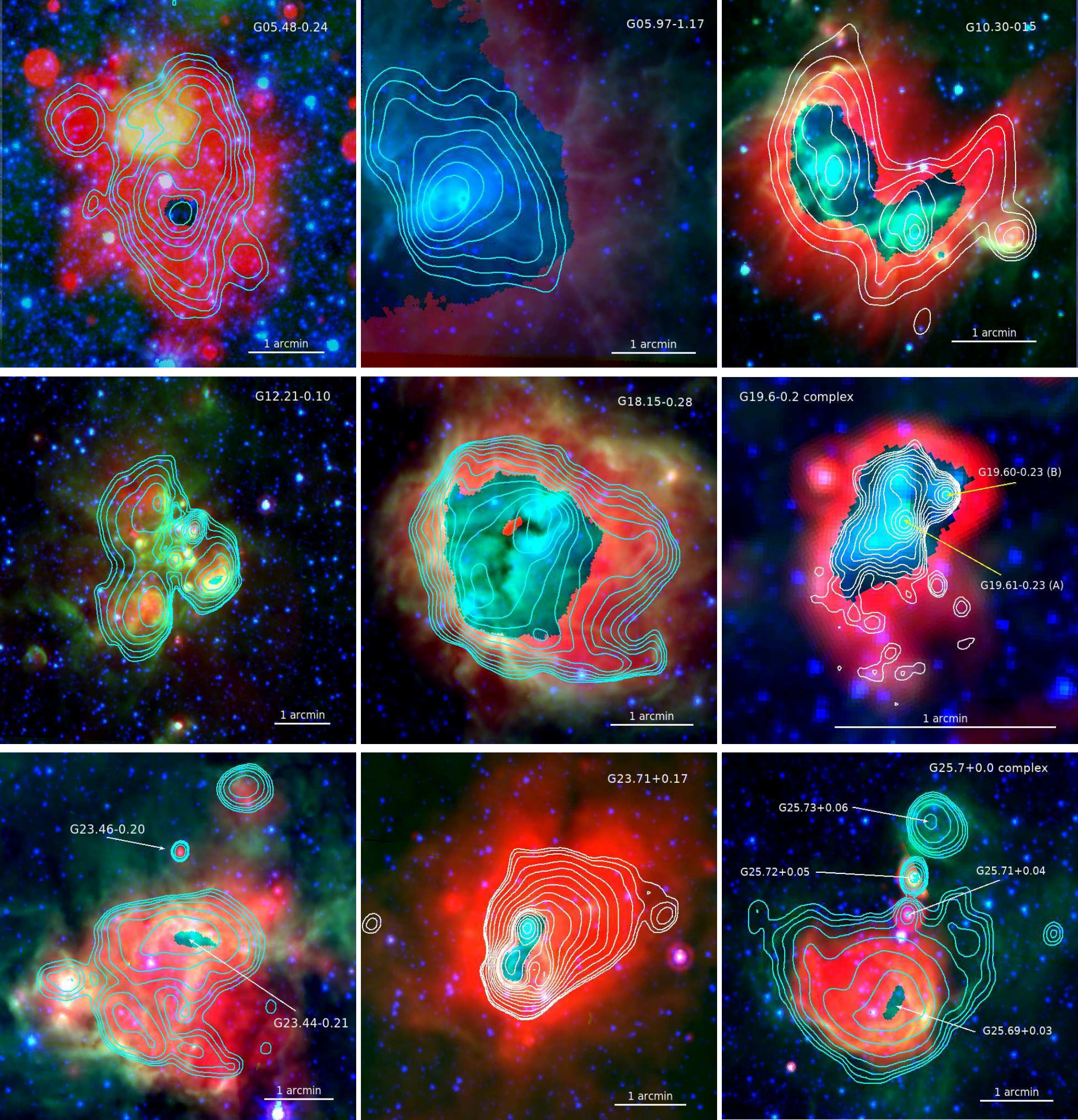}

  \caption{RGB images (Red = 24~\micron, Green = 8.0~\micron, and Blue = 4.5~\micron ) of the \uchiies~ with MIPS data available. Contours at 3.6~cm are from VLA D configuration (see Table~\ref{tab:tab2}), except for G19.60$-$0.23 (BD), G31.39-0.25 (BCD), G37.87+0.40 (BCD), and G60.88$-$0.13 (BC) where MRIs are superimposed (see Table~\ref{tab:tab3}). All these sources are catalogued as \uchiies~ except G35.58--0.03 and G48.61$+$0.02. These images show that in these regions, the red emission (dust) matches very well with the whole radio continuum emission, the green emission (PAH$'$s) delineates the radio continuum cometary arcs, and the \uchii~ position is saturated at 24~\micron . See text in \S~\ref{indSources} for individual details.}
  \label{fig:fig3a}
\end{figure*}

  \setcounter{figure}{2}
\begin{figure*}
    \includegraphics[width=0.94\textwidth]{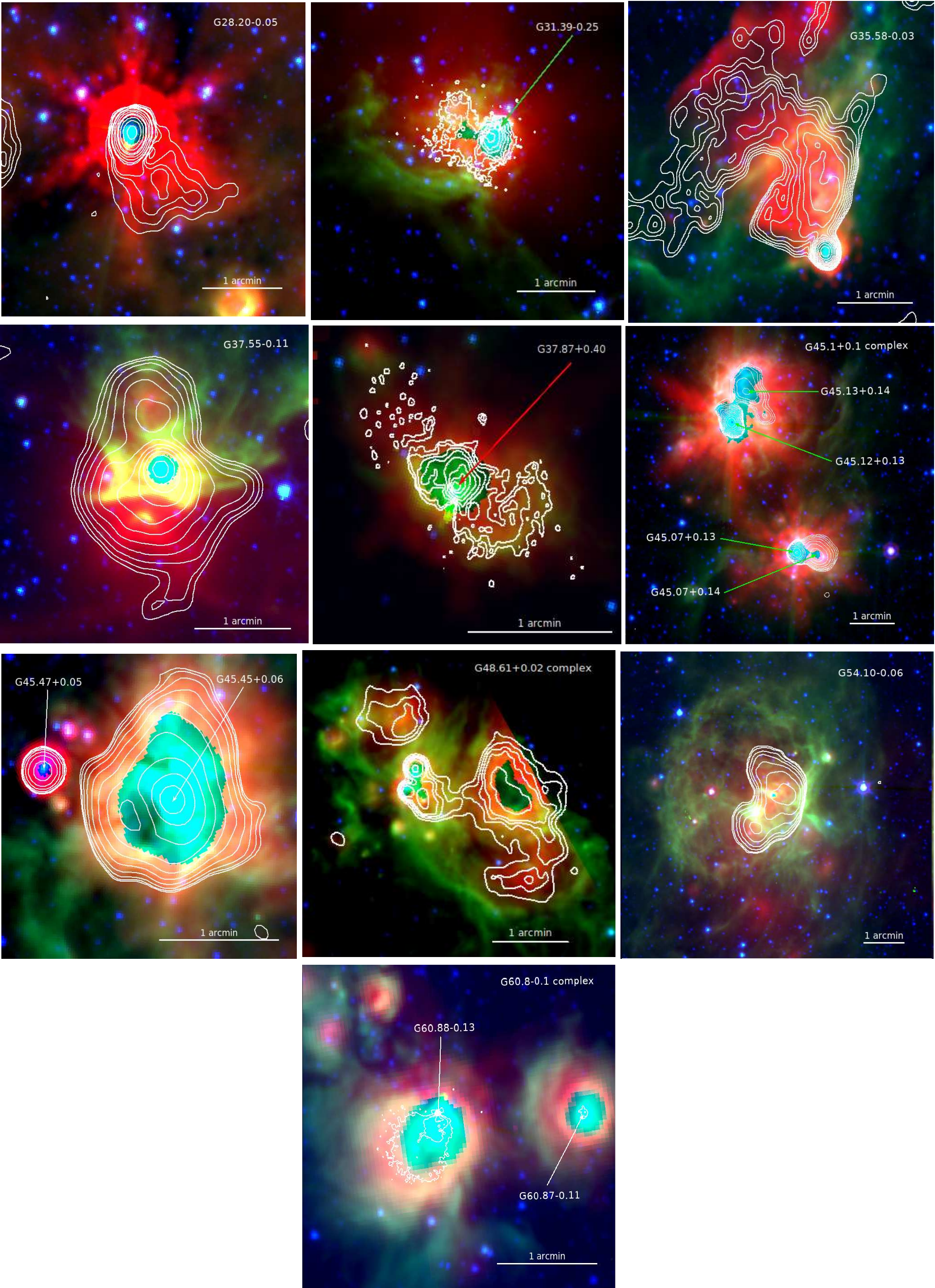}

  \caption{\textit{continued.}}
  \label{fig:fig3b}
\end{figure*}

The literature shows that all the \uchiies~ are related with star--forming regions, with YSOs and other related phenomena such as maser emission. Also, many of them were catalogued by \citet{Ch06, Ch07} and \citet{Simp12}. The last columns in Table~\ref{tab:tab4} summarize a search of the sources in our catalogue of 28 \uchiies~ in other IR surveys. Several sources have not had an \uchiie~ classification before. 

Combining the \textit{Spitzer} view with our radio continuum results, it can be seen that the \uchiies~ of our catalogue are arc--min scale star--forming regions. In some cases they are part of larger (degree size) structures, and composed of a collection of arc--min scale ionized gas bubbles similar to those found by \citet{Ch06} and \citet{Simp12}. Alternatively, they may be compact \hii~ regions  \citep{PH08} with well defined cometary arcs, where low--mass and high--mass YSOs, ionized gas, star clusters, dust and PAH$'$s coexist. This result is more realistic than the idea of a single ionizing source producing the EE involving or not a density gradient from a champagne flow model. In addition, it is very likely that the difference between an \uchiie~ and a compact \hii ~region is more an effect of scale--size of the emission than an evolutionary trend from the \uchii~ region defined by \citet{WC89} and \citet{K94}.

\section{Comments on Individual Sources}
\label{indSources}

Based on the analysis of radio continuum and IR images (Fig.~\ref{fig:fig1a}, \ref{fig:fig2a} and \ref{fig:fig3a}), a confirmation, addition or elimination to a final catalogue of \uchiies~ was performed. As a result, 68\% of the original sample (19/28 regions) were confirmed to be \uchiies, 7\% (2/28) were re--assigned a new RC--peak or ultra--compact source(s) with the EE, and 25\% (7/28) were not defined as \uchiies. A total of 21 \uchiies~ are reported in Table~\ref{tab:tab5}. Below, we provide the IRAC--MIPS RGB images in Fig.~\ref{fig:fig3a}, and comment on individual regions.

\begin{table*}
\caption{The Final Catalogue of \uchiie}
\label{tab:tab5}
\begin{threeparttable}
\begin{tabular}{clcccccc}

\hline

\multicolumn{1}{c}{\uchiie} & \multicolumn{1}{c}{IRAS}  & \multicolumn{1}{c}{R. A.} & \multicolumn{1}{c}{Dec.} & \multicolumn{1}{c}{Distance\tnote{a}}  & \multicolumn{1}{c}{Size} & \multicolumn{1}{c}{Spitzer} & \multicolumn{1}{c}{RC} \\

\multicolumn{1}{c}{region} & \multicolumn{1}{c}{Source}  & \multicolumn{1}{c}{(J2000)} & \multicolumn{1}{c}{(J2000)} & \multicolumn{1}{c}{(kpc)}  & \multicolumn{1}{c}{approx} & \multicolumn{1}{c}{Bubble\tnote{b}} & \multicolumn{1}{c}{Peaks} \\
 \hline 
 
G05.48--0.24                  & 17559--2420 & 17 59 02.9 & --24 20 54.5 &14.3$^{(1)}$ & $3.0' \times 2.0'$    & Y & 5  \\
G05.97--1.17                  & 18006--2422 & 18 03 40.5 & --24 22 44.7 & 2.7$^{(2)}$ & $3.5' \times 2.0'$  & Y & 1  \\
G10.30--0.15                 & 18060--2005 & 18 08 56.1 & --20 05 53.4 & 6.0$^{(3)}$ & $4.0' \times 1.5'$  & Y & 5  \\
G12.21--0.10\tnote{c}  & 18097--1825A& 18 12 39.7 & --18 24 20.1 &13.5$^{(2)}$ & $4.0' \times 3.0'$    & Y & 5  \\
G18.15--0.28                 & 18222--1321 & 18 25 01.2 & --13 15 40.0 & 4.2$^{(3)}$ & $2.0' \times 4.0'$    & Y & 2  \\
G19.60--0.23                 & 18248--1158 & 18 27 37.7 & --11 56 42.0 & 3.5$^{(2)}$ & $0.8' \times 2.0'$ & Y & 5  \\
G23.44--0.21                 & 18319--0834 & 18 34 44.4 & --08 32 22.5 & 9.0$^{(1)}$ & $2.0' \times 2.0'$    & Y & 1  \\
G23.71+0.17                  & 18311--0809 & 18 33 53.5 & --08 07 14.2 & 8.9$^{(2)}$ & $2.0' \times 2.0'$    & Y & 1  \\
G25.69+0.03          & 18353--0628  & 18 38 05.1 & --06 25 31.9 & 9.3$^{(2)}$ & $3.0' \times 4.0'$    & Y & 2  \\
G28.20--0.05                 & 18402--0417 & 18 42 58.2 & --04 14 05.0 & 9.1$^{(3)}$ & $1.5' \times 1.5'$  & Y & 1  \\
G31.39--0.25                 & 18469--0132 & 18 49 32.9 & --01 29 03.7 & 8.9$^{(5)}$ & $2.5' \times 1.5'$& Y & 2  \\
G35.20--1.74                 & 18592+0108  & 19 01 46.9 & +01 13 08.0  & 3.3$^{(4)}$ & $2.0' \times 2.0'$    & P & 1  \\
G37.55--0.11                 & 18577+0358  & 19 00 16.0 & +04 03 10.3  & 9.9$^{(2)}$ & $2.0' \times 2.0'$    & Y & 2  \\
G37.87+0.40                  & 18593+0408  & 19 01 53.6 & +04 12 48.9  & 9.3$^{(4)}$ & $2.0' \times 0.5'$  & Y & 1  \\
G45.07+0.13                  & 19110+1045  & 19 13 22.0 & +10 50 53.9  & 6.0$^{(4)}$ & $1.0' \times 1.0'$    & Y & 1  \\
G45.12+0.13                  & 19111+1048  & 19 13 27.8 & +10 53 36.7  & 6.0$^{(4)}$ & $1.5' \times 1.5'$& Y & 1  \\
G45.45+0.06                  & 19120+1103  & 19 14 21.3 & +11 09 12.9  & 6.0$^{(4)}$ & $2.0' \times 2.0'$    & Y & 1  \\
G54.10--0.06                 & 19294+1836  & 19 31 43.1 & +18 42 52.0  & 7.9$^{(1)}$ & $3.0' \times 2.0'$    & Y & 2  \\
G60.88--0.13\tnote{d}  & 19442+2427  & 19 46 20.1 & +24 35 29.4  & 2.2$^{(4)}$ & $1.0' \times 3.0'$    & Y & 1  \\
G77.96--0.01                 & 20277+3851  & 20 29 36.7 & +39 01 21.9  & 4.4$^{(5)}$ & $2.0' \times 3.0'$    & P & 2  \\
G111.28--0.66                & 23138+5945  & 23 16 03.9 & +60 02 00.8  & 2.5$^{(3)}$ & $2.5' \times 2.5'$& P & 2  \\
\hline
\end{tabular}

\begin{tablenotes}
\item[a] From: (1) \citet{WC89}, (2) \citet{CWC90}, (3) \citet{K94}, (4) \citet{Ar02} and (5) \citet{K99}.

\item[b] In present work $Y$ = Yes and $P$ = Probably, based on the availability of \textit{Spitzer} data. However, for those marked with $P$ and without MIPS data, the 2MASS K$_{\rm s}$ image suggests that gas emission matches the morphology of radio continuum data (see Fig.~\ref{fig:fig4}). 

\item[c] This source was studied in detail and confirmed to host a hot core by \citet{F18}.

\item[d] The source G60.87--0.11, located $\sim$1.5\arcmin ~to the West of this source, is proposed as a new \uchiie, but more data are needed to confirm this.

\end{tablenotes}
\end{threeparttable}
\end{table*}

\subsection {Sources confirmed as \texorpdfstring{UC~H\,{\sevenrm II} + EE}~ (19 regions)}   
\label{confirm}

\begin{description}

\item G05.48--0.24: This source was already classified by \citet{Ko96} as an \uchiie ~. They presented VLA AnB and BnC radio continuum data as well as VLA D data at 21 cm, indicating that the \uchiir ~is immersed in extended arc--min scale diffuse emission. In addition, they also presented HI and $^{13}$CO line data. Our VLA--D image (Fig.~\ref{fig:fig1a}) confirms that this source can be classified as \uchiie. The MIPS image in Fig.~\ref{fig:fig3a} shows saturation on the \uchii~ position and the EE matches very well with the MIPS emission.

\item G05.97--1.17: \citet{Steck98} argued that the nature of this UC source should not be considered as an \uchiir, ionized by an embedded B0 star, but rather as a young star surrounded by a circumstellar disk that is being photo--evaporated by Her~36. On the other hand \citet{KK01} described this source as a single compact component, probably Her~36, surrounded by extended emission of 14\arcmin $\times$ 10.7\arcmin size. Furthermore, it can be seen from Fig.~\ref{fig:fig3a} that all the EE presented in Fig.~\ref{fig:fig1a} is located within the region where the MIPS emission is saturated, which supports the description given by \citet{KK01}. In addition, since the EE resembles the \textit{Spitzer} IR bubbles, we decided to include this source in our catalogue.

\item [G10.30--0.15:] The image in Fig.~\ref{fig:fig3a} shows saturated MIPS emission on the UC component described by \citet{KK01, KK02} confirming its nature as \uchiie~region. MIPS 24~\micron ~emission coincides with the VLA emission presented in Fig.~\ref{fig:fig1a}.

\item [G12.21--0.10:] This source was also included in studies by \citet{KK01,KK03}. They observe VLA DnC 21 cm on radio--continuum and radio--recombination lines, as well as with the NRA0 12 m CO and CS molecular tracers. \citet{F18}, confirm the presence of a new hot molecular core and provide its characterization. Its RGB image in Fig.~\ref{fig:fig3a} shows a collection of compact \hii~ regions with cometary arcs well defined by the 8.0~\micron ~emission.

\item [G18.15--0.28:] Before this work, no VLA low angular resolution images of this object were published. \citet{WC89} and \citet{K94} classified this source as cometary. Our 3.6~cm VLA--D image shows extended emission around the UC component labeled with an arrow in Fig.~\ref{fig:fig1a}. The MIPS emission is saturated at arc-min scales, suggesting that several \uchiirs ~might be in this zone.

\item [G19.60--0.23:] This source was studied in detail by \citet{Ga98} but never classified as \uchiie. Our VLA--D image at 3.6~cm (Fig.~\ref{fig:fig1a}), and the MRI combining this data with the 3.6~cm VLA-B from \citet{K94} confirm this classification. The MIPS image in Fig.~\ref{fig:fig3a} shows saturation in the \uchii~ regions G19.61--0.23 and G19.60--0.23 as well as on the EE ($\sim$30\arcsec ). This image not only confirms the nature as an \uchiie~ region of the G19.60--0.2 complex, but also that the EE defines an IR bubble.

\item [G23.71$+$0.17:] Fig.~\ref{fig:fig3a} shows MIPS saturated emission at the \uchii~ position reported by \citet{KK01} and at the SE component marginally detected in the \citet{KK01} image. This image suggests that at least two \uchii~ regions might be present.

\item [G28.20--0.05:] It was classified as hypercompact \hii ~ by \citet[][]{Se08}. A visual inspection on the 8~\micron~ image at large scale suggests this object as an \uchiie~ into an Infrared Dark Cloud (IRDC), where IRAC emission does not trace the radio continuum emission but covers all the region in a continuous structure of 15$'$ in size. On the other hand, the MIPS image shown in Fig.~\ref{fig:fig3a} clearly shows that the 24~\micron ~emission dominates and coincides with the radio continuum extended emission, and it is saturated at the UC position.  

\item [G31.39--0.25:] \citet{K99} suggested this source as an \uchiie~and reported two sources, G31.396-0.257 (marked with an arrow in Fig.~\ref{fig:fig2a}) and a component slightly to the North, G31.397-0.257. Besides, our map also shows a third radio continuum peak located at 18h:49m:33.4s,-01d:29m:11.4s which corresponds to the YSO MSX6CG031.3948-00.2585 reported by \citet{Ur09}. Part of this bubble is delineated by the southern arc of about 2\arcmin~ in length shown in 8~\micron~ emission (Fig.~\ref{fig:fig3a}). 

\item [G35.20--1.74:] The new VLA images at 3.6~cm taken with configurations D and C are shown in Figs.~\ref{fig:fig1a} and \ref{fig:fig2a}, respectively. Since a density gradient n$_{\rm e}$ $\propto$ r$^{-2}$ was reported by \citet{Fr00} and the radio continuum morphology resembles a champagne flow, this source could be considered as a candidate to test the \citet{KK01, KK03} model. Unfortunately, no {\it Spitzer} data are available. Nevertheless, the K$_{\rm s}$ 2MASS image shows diffuse gas that matches with the radio continuum extended emission (see Fig.~\ref{fig:fig4}). On the other hand, if the K$_{\rm s}$ 2MASS emission turned out to be similar to that observed at {\it Spitzer} wavelengths, then this EE could be an IR--bubble in the style of Churchwell et al. However, further observations would be required to confirm this hypothesis.

\begin{figure*}
\centering
    \includegraphics[width=\textwidth]{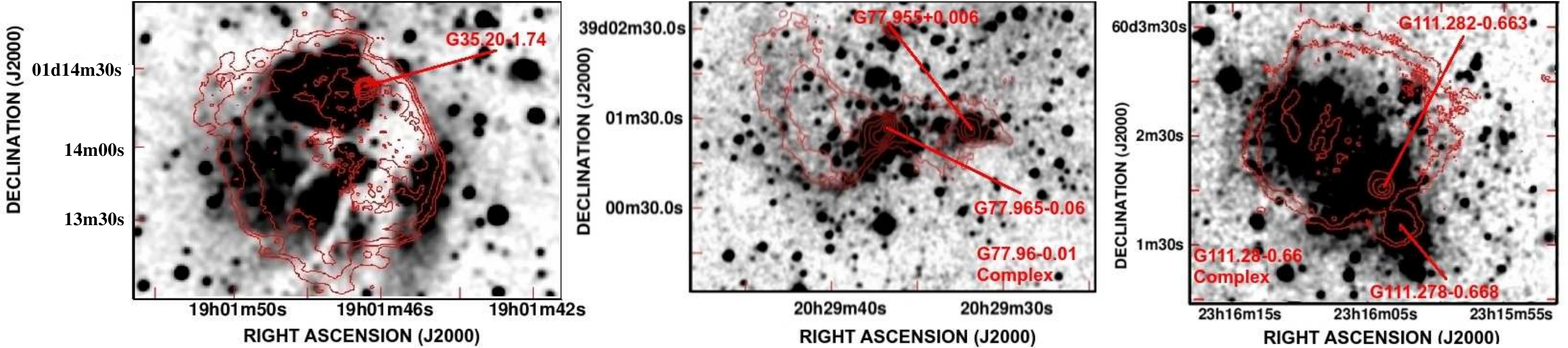}

  \caption{2MASS K$_{\rm s}$ band images for the three sources classified as P=Probably in column 7 of Table~\ref{tab:tab5}. Contours at 3.6~cm are from multi--resolution maps (Fig.~\ref{fig:fig2a}). These regions do not have MIPS data.}
  \label{fig:fig4}
\end{figure*}

\item [G37.55--0.11:] This source was catalogued as \uchiie~ by \citet{KK01}. The IRAC bands image shows a remarkable lack of IRAC emission at the South part of the \uchii~ region, i.e. there is no IRAC emission tracing this part of the EE. On the other hand, MIPS emission is observed to trace very well the ionized gas (Fig.~\ref{fig:fig3a}). This behavior can be explained in terms of colder dust present in this region of the \uchiie~ that is clearly bounded to the north by emission by PAH$'$s (yellow and green colours). Furthermore, the \uchii~ position is saturated. The MIPS emission is a good tracer of \uchii~ regions and extended ionized gas (Fig.~\ref{fig:fig1a}).

\item [G37.87$+$0.40:] This source has bipolar morphology on the VLA--D image at 3.6~cm \citep{K99}. However, on the configuration C and multi--resolution image (Fig.~\ref{fig:fig2a}), the NE lobe is weak and almost disappears while the SW lobe emission is more intense. Although the bipolar morphology of this source could be due to a bipolar flow, there is not direct evidence of its existence. An analysis of the CO line emission could shed light on the origin of the bipolar morphology. Based on our morphological study and the fact that both UC and EE in central contours coincide with the saturated region in 24~\micron~ MIPS emission (see Fig~\ref{fig:fig3a}), we suggest that G37.87$+$0.40 is an \uchiie.

\item [G45.07$+$0.13 and G45.12$+$0.13:] These two objects are located in a 5\arcmin ~field, and both present extended emission as shown in Fig.~\ref{fig:fig1a}. The MIPS image shows saturation on the \uchiir~ and the EE confirming that this complex has at least 4 \uchii~ regions: G45.12$+$0.13 and G45.13$+$0.14 to the North, and G45.07$+$0.13 and G45.07$+$0.14to the South.

\item [G45.45$+$0.06 and G45.47$+$0.05:] In the VLA--D image at 3.6~cm, two \uchii~ regions are observed in the field. One corresponds to G45.45$+$0.06 (marked with an arrow in Fig.~\ref{fig:fig1a}), and the other, more compact, to the East is G45.47$+$0.05. In the NRAO VLA Sky Survey image at 21~cm \citep{Co98}, both \uchii~ regions are surrounded by the same radio continuum emission. Our image, shown in Fig.~\ref{fig:fig1a}, confirms that G45.45$+$0.06 can be classified as \uchiie. The MIPS image in Fig.~\ref{fig:fig3a} confirms this result indicating that for G45.45$+$0.06 more than one \uchii~ region can be included, since the shape of saturation in not circular and some point-like sources can be seen at 4.5~\micron.

\item [G54.10--0.06:] Before this work, no low resolution VLA images of this source were reported. Our VLA--D image at 3.6~cm shows extended emission surrounding two radio continuum peaks, the most prominent marked with an arrow in Fig.~\ref{fig:fig1a} (G54.10--0.06), and a second source to the SE. The MIPS emission is saturated on the \uchii~ position and the EE resembles 2 bubbles with PAH$'$s emitting filaments or a PDR that clearly separates them.

\item [G60.88--0.13:] This source is the only one with \textit{Spitzer} emission and radio continuum morphology that may be explained under the scenario of a central ionizing source, producing an ionization front propagated in a density gradient medium, and producing a champagne flow \citep{KK01,KK03,Fr90}. Further molecular observations are needed to confirm if this model applies. However, since the \textit{Spitzer} emission coincides completely with the radio continuum emission, the associated EE has the bubble morphology discussed by \citet{Ch06} and \citet{Simp12}. 


\item [G77.96--0.01:] In Fig.~\ref{fig:fig2a}, this source presents two \uchii~ regions; one at the centre (G77.965--0.006), and the other $\sim$ 1\arcmin ~to the West (G77.955$+$0.006). Both are located inside the EE in a 'double--lobe' structure. On the configuration C image (Fig.~\ref{fig:fig2a} left panel), the emission in the West lobe from G77.965--0.006 toward G77.955$+$0.006 tends to disappear but surrounds G77.955$+$0.006. Unfortunately, no \textit{Spitzer} data are available for this source, but the 2MASS K$_{\rm S}$ image shows gas and several point sources see Fig.~\ref{fig:fig4}. The observed gas matches with the radio continuum extended emission but it is very weak in the zone where the two lobes become very faint. Both lobes may be part of different bubbles, with G77.965--0.006 being the object that can be classified as \uchiie, but better studies, kinematics, and IR images are needed to confirm this.

\item [G111.28--0.66:] This object was catalogued as \uchiie~ by \citet{K94}. Two \uchii~ regions are clearly observed in Fig.~\ref{fig:fig2a}: G111.278--0.668 and G111.282--0.663, this last identified with an arrow. Unfortunately, the MIPS image is not available. However, the K$_s$ 2MASS nebulosity resembles the extended emission observed in the SW--NE direction (see Fig.~\ref{fig:fig4}). Still, mid IR images are needed to confirm if the EE can be considered as an IR bubble, but there is no reason to discard this object as an \uchiie, following \citet{K94} and \citet{K99}.

\end{description}

\subsection{Sources with \texorpdfstring{ UC~H\,{\sevenrm II}}~ region re--assigned (2 regions)}
\label{re_assign}

\begin{description}

\item [G23.46--0.20:] \citet{KK01} showed a multi--peak image where the main source is marked with a solid arrow in Fig.~\ref{fig:fig1a}. Our VLA--D, 3.6~cm image shows 3 independent sources discarding the possible association  of G23.46--0.20 with the EE. Although G23.46--0.20 can be discarded as an \uchiie, the extended emission is more related to the source labelled as A by \citet{KK01} and shown with a dashed arrow in Fig.~\ref{fig:fig1a}.  This new source is designated as G23.44--0.21 in this work (see Table~\ref{tab:tab5}). This re--assignment is also confirmed with the MIPS image (Fig.~\ref{fig:fig3a}), where G23.46--0.20 appears as a single \uchii~ region while G23.44--0.21 appears saturated, and the EE coincides with the dust. The shape of saturation appears elongated in the E--W direction and two--three point--like counterpart sources are seen in 4.5~\micron. Since one or more of these sources might be \uchiirs ~we decided to keep it as an \uchiie~ rather than consider it as one compact \hii ~region. Observations with higher angular resolution that resolve out the EE could single out the more compact emission from the \uchiirs.

\item [G25.72$+$0.05:] \citet{KK01} do not classify this source as \uchiie~ because the LSR velocities, which were obtained from radio--recombination lines, were different for the UC emission and the EE. Our VLA--D 3.6~cm image has better resolution than their data and three aligned but independent point--sources are clearly observed (see Fig.~\ref{fig:fig1a}). The EE includes more to G25.71$+$0.04 (dashed arrow in Fig.~\ref{fig:fig1a}) than to G25.72$+$0.05 (solid arrow in the same figure). The source G25.73$+$0.06, to the North of G25.72$+$0.05 does not have MIPS counterpart data, thus we do not know if its 24~\micron ~emission is saturated, but it could be another \hii~ region, maybe compact. Nevertheless, this source appears not to be related to the EE. The MIPS image of the G25.7$+$0.0 complex in Fig.~\ref{fig:fig3a} confirms these results and strongly suggests that the EE is not only associated with G25.71$+$0.04 but mostly related to G25.69$+$0.03. As in the case of G23.44--0.21, G25.69$+$0.03 might contain a set of \uchiirs~in the NW--SE direction as shown by the saturated region.  

\end{description}

\subsection{Sources discarded as \texorpdfstring{UC~H\,{\sevenrm II} + EE}~ (7 regions)}
\label{discarded}

\begin{description}

\item [G33.13--0.09:] was classified as an 'unlikely' \uchiie~ by \citet{K99}. Our VLA MRI confirms this interpretation showing that the \uchii~ region and the EE are independent of one another (see Fig.~\ref{fig:fig2a}).

\item [G35.58--0.03:] was classified as an 'unlikely' \uchiie~ by \citet{K99}. Our VLA MRI (see Fig.~\ref{fig:fig2a}) confirms this assumption showing that the EE is more related to another structure, like a complex elongated SE$-$NW structure. The MIPS image (Fig.~\ref{fig:fig3a}) shows saturation at the position of the \uchii~ region and the EE coincides with two regions separated by a PAH$'$s filament or a PDR (green and yellow emission).

\item [G43.24--0.04:] Our VLA MRI is presented in Fig.~\ref{fig:fig2a}. It is totally discarded as \uchiie~ as was suggested by \citet{K99}.

\item [G48.61$+$0.02:] \citet{K94} report three Ultra--compact components in this field; G48.606$+$0.023 (to the SE), G48.606$+$0.024 (at the middle), and G48.609$+$0.027 (to the North). These sources are marked with arrows in Fig.~\ref{fig:fig2a}.  Although \citet{K99} classify this region as a 'possible' \uchiie, our images discard this idea. In Fig.~\ref{fig:fig3a}, two saturated zones are seen. MIPS saturated zone to the East coincides with the three sources reported by \citet{K94}, and a star cluster is known \citep{Mo13}, [MCM2005b] 21. Saturated zone to the West coincides with the radio source GB6 B1918+1349 \citep{Gr96} and about 10 IR sources in the 2MASS catalogue \citep{Sk06} in a radius of about 20\arcsec.

\item [G78.44$+$2.66:] \citet{K99} argued for an 'unlikely' connection between the EE and the \uchii~ region. Our images confirm this idea (see Fig.~\ref{fig:fig2a}) and no \textit{Spitzer} data are available.

\item [G106.80$+$5.31:] This field contains three Ultra--compact components \citep{K99} marked with arrows in Fig.~\ref{fig:fig2d}; G106.795$+$5.311 (to the southwest), G106.797$+$5.312 (in the middle), and G106.800$+$5.316 (to the north). \citet{K99} argued a connection was unlikely between the EE and the \uchii~ region. Our images (see Fig.~\ref{fig:fig2a}) confirm this idea.  No \textit{Spitzer} image is available and we cannot classify it as a bubble.

\item [G111.61$+$0.37:] This case is quite similar to G106.80$+$5.31 and G78.44$+$2.66. We confirm the result of \citet{K99} of not classifying this source as \uchiie, and no \textit{Spitzer} image is available. 

\end{description}

\section{Energetics in \texorpdfstring{UC~H\,{\sevenrm II} +}~ EE}
\label{energy}

Following \citet{K99}, a way to quantify the energetics of an \uchiie~ is comparing the total ionizing flux of the exciting star calculated from radio continuum and Far Infrared (FIR) observations. The ionizing photon rate, $N'_c$, is estimated from the 3.6~cm observations in the D configuration using equation 1 of \citet{K94}, 

\begin{equation}
\left(\frac{N'_c}{s^{-1}}\right)~\ge~8.04\times10^{46}\left(\frac{T_e}{K}\right)^{-0.85}\left(\frac{r}{pc}\right)^3\left(\frac{n_e}{cm^{-3}}\right)^2,
\end{equation}

\noindent considering a homogeneous sphere of radius $r$, electron density $n_e$, and electronic temperature $T_e$.




On the other hand, the total luminosity measured via IRAS fluxes can be converted into an ionizing photon rate, N$^*_c$, using model stellar atmospheres \citep[e.g.,][]{Pa73, Ca86, WC89,   K94}. N$'_c$ and N$^*_c$ are considered limit values because an ionization--bounded and dust--free nebula is assumed; N$'_c$ represents a lower limit to the spectral type, while  N$^*_c$ an upper limit \citep{K94, F09a}.

Thus, we are interested in estimate the fraction of UV photons absorbed by dust --and hence without causing ionization--, f$_d$, defined as \citep{K94} 

\begin{equation}
f_d = 1-\frac{N'_c}{N^*_c},
\end{equation}

\noindent which we also assumed to produce all the IRAS FIR luminosity by dust within the \uchiie. 



As mentioned in \S~\ref{intro}, the infrared excess is quantified by f$_d$. Values of f$_d$ $\sim$ 1 indicate a large IR--excess. Values of 0.42~$<$~f$_d$~$<$~0.99 where found by the surveys of \citet{WC89} and \citet{K94}.

A comparison between f$_d$ (configuration D; \citealt{K99} and this paper) and f$_d$ (configuration B; \citealt{WC89} and \citealt{K94}) was performed. We use configuration D data because this configuration has a MRS of up to 3\arcmin, and f$_d$ (configuration D) is slightly smaller than f$_d$ (configuration C). Both configuration B and configuration D values are reported in Table~\ref{tab:tab6}. In order to determine the respective N$'_c$, we use the flux taken from a box that covers all the radio continuum emission observed.  

\begin{table*}
\caption{Results on the energetics for the Initial Sample}
\label{tab:tab6}
\begin{threeparttable}
\begin{tabular}{clccccccc}

\hline

\multicolumn{1}{c}{\uchii} & \multicolumn{1}{c}{L$_{\rm IRAS}$} & \multicolumn{1}{c}{Log N*$_{\rm c}$} & \multicolumn{1}{c}{Spectral Type} & \multicolumn{1}{c}{VLA} & \multicolumn{1}{c}{Log N$'_{\rm c}$} & \multicolumn{1}{c}{Spectral Type} & \multicolumn{1}{c}{f$_{\rm d}$(C,D) / f$_{\rm d}$(B max)\tnote{a}} & \multicolumn{1}{c}{EE level\tnote{b}} \\

\multicolumn{1}{c}{region} & \multicolumn{1}{c}{(10$^4$ L$_{\odot}$)} & \multicolumn{1}{c}{(s$^{-1}$)} & \multicolumn{1}{c}{(IRAS)} & \multicolumn{1}{c}{(configuration)} & \multicolumn{1}{c}{(s$^{-1}$)} & \multicolumn{1}{c}{(radio)} &  & \multicolumn{1}{c}{Explain} \\
\hline

G05.97--1.17 & 15.5
& 48.82 & O6.5 & D & 48.75 & O6.5  & 0.16 / 0.90$^{\rm (1)}$ & 1 \\ 
G19.60--0.23 & 14.0 & 48.82 & O6.5 & D & 48.70 & O7    & 0.24 / 0.50$^{\rm (2)}$ & 1 \\
G33.13--0.09 & 10.3 & 48.62 & O7   & D & 48.42 & O8    & 0.37 / 0.48$^{\rm (2)}$ & 1 \\
	         &       &       &      & C & 48.41 & O8    & 0.38 / \,\,\, $-$ \,\,&   \\
G35.20--1.74 & 26.9 & 49.08 & O6   & D & 49.07 & O6    & 0.03 / 0.79$^{\rm (2)}$ & 1 \\   
             &       &       &      & C & 49.05 & O6    & 0.06 / \,\,\, $-$ \,\,&   \\
G35.58--0.03 &  4.7 & 48.08 & O9   & D & 48.02 & O9    & 0.13 / 0.90$^{\rm (2)}$ & 1 \\ 
	         &       &       &      & C & 47.93 & O9.5  & 0.29 / \,\,\, $-$ \,\,&   \\ 
G37.55--0.11 & 28.1 & 49.08 & O6   & D & 48.94 & O6.5  & 0.13 / 0.90$^{\rm (1)}$ & 1 \\     
G37.87+0.40  & 65.1 & 49.62 & O5   & D & 48.98 & O6.5  & 0.19 / 0.44$^{\rm (2)}$ & 1 \\
	         &       &       &      & C & 48.96 & O6.5  & 0.20 / \,\,\, $-$ \,\,&   \\  
G43.24--0.04 & 31.7 & 49.08 & O6   & D & 48.98 & O6.5  & 0.21 / 0.81$^{\rm (2)}$ & 1 \\
             &       &       &      & C & 48.96 & O6.5  & 0.24 / \,\,\, $-$ \,\,&   \\
G45.45+0.06  & 49.2 & 49.36 & O5.5 & D & 49.19 & O6    & 0.33 / 0.96$^{\rm (2)}$ & 1 \\
G111.28--0.66 & 2.7 & 47.36 & B0   & D & 47.25 & B0    & 0.22 / 0.87$^{\rm (2)}$ & 1 \\
	         &       &       &      & C & 47.17 & B0    & 0.35 / \,\,\, $-$ \,\,&   \\
\hline
\hline
G05.48--0.24 & 66.5 & 49.62 & O5   & D & 49.36 & O5.5 & 0.45 / 0.95$^{\rm (1)}$ & 2 \\
G10.30--0.15 & 65.7 & 49.62 & O5   & D & 49.30 & O5.5 & 0.52 / 0.96$^{\rm (1)}$ & 2 \\
G18.15--0.28 & 21.6 & 49.08 & O6   & D & 48.85 & O6.5 & 0.42 / 0.99$^{\rm (2)}$ & 2 \\
G23.71+0.17  & 34.2 & 49.36 & O5.5 & D & 49.05 & O6   & 0.51 / 0.95$^{\rm (1)}$ & 2 \\
G25.69+0.03  & 22.1 & 49.08 & O6   & D & 48.75 & O6.5 & 0.54 / 0.99$^{\rm (1)}$ & 2 \\
G28.20--0.05 & 20.5 & 49.08 & O6   & D & 48.58 & O7   & 0.69 / 0.93$^{\rm (2)}$ & 2 \\
G48.61+0.02  & 74.5 & 49.62 & O5   & D & 49.17 & O6   & 0.64 / 0.99$^{\rm (2)}$ & 2 \\
	         &       &       &      & C & 48.87 & O6.5 & 0.82 / \,\,\, $-$ \,\,&   \\
G54.10--0.06 & 14.9 & 48.82 & O6.5 & D & 48.49 & O7.5 & 0.53 / 0.99$^{\rm (1)}$ & 2 \\
G77.96--0.01 & 10.2 & 48.62 & O7   & D & 48.24 & O8.5 & 0.58 / 0.96$^{\rm (2)}$ & 2 \\
	         &       &       &      & C & 48.04 & O9   & 0.74 / \,\,\, $-$ \,\,&   \\
G111.61+0.37 & 16.6 & 48.82 & O6.5 & D & 48.40 & O8   & 0.62 / 0.90$^{\rm (2)}$ & 2 \\
	         &       &       &      & C & 48.38 & O8   & 0.64 / \,\,\, $-$ \,\,&   \\
\hline	     
\hline	     
G12.21--0.10 & 86.8 & 49.93 & O4   & D & 49.36 & O5.5 & 0.73 / 0.94$^{\rm (1)}$ & 3 \\	     
G23.44--0.21 & 73.9 & 49.62 & O5   & D & 48.95 & O6.5 & 0.79 / 0.99$^{\rm (1)}$ & 3 \\
G31.39--0.25 & 38.9 & 49.36 & O5.5 & D & 48.69 & O7   & 0.79 / 0.99$^{\rm (2)}$ & 3 \\
	         &       &       &      & C & 48.69 & O7   & 0.79 / \,\,\, $-$ \,\,&   \\
G45.07+0.13  & 49.3 & 49.36 & O5.5 & D & 48.40 & O8   & 0.89 / 0.94$^{\rm (2)}$ & 3 \\
G45.12+0.13  & 62.6 & 49.62 & O5   & D & 48.79 & O6.5 & 0.85 / 0.71$^{\rm (2)}$ & 3 \\ 
G60.88--0.13 &  4.3 & 48.08 & O9   & C & 47.35 & B0   & 0.81 / 0.97$^{\rm (2)}$ & 3 \\
G78.44+2.66  &  6.1 & 48.35 & O8   & D & 46.86 & B0.5 & 0.90 / 0.99$^{\rm (2)}$ & 3 \\
	         &       &       &      & C & 46.86 & B0.5 & 0.97 / \,\,\, $-$ \,\,&   \\
G106.80+5.31 &  2.3 & 47.36 & B0   & D & 45.19 & B1   & 0.90 / 0.99$^{\rm (2)}$ & 3 \\
\hline

\end{tabular}

\begin{tablenotes}
\item[a] Values of f$_d$ based on VLA C or D configuration to compare with f$_d$ based on VLA B configuration data. For the latter, upper numbers denote values from: (1) \citet{WC89} and (2) \citet{K94}.

\item[b] Different levels at which EE reduces the values of ~f$_d$ explaining the \textit{IR--Excess}: 1.- Low, 2.- Intermediate, 3.- High. (see \S~\ref{energy} for details).

\end{tablenotes}
\end{threeparttable}
\end{table*}

\begin{table*}
\caption{Energetics Results for the Final Catalogue}
\label{tab:tab7}
\begin{threeparttable}
\begin{tabular}{clccccccc}

\hline

\uchii & L$_{\rm IRAS}$\tnote{a} & Log N*$_{\rm c}$\tnote{a} & Spectral Type\tnote{a} & VLA\tnote{b} & Log N$'_{\rm c}$\tnote{b} & Spectral Type\tnote{b} & f$_{\rm d}$(B max)\tnote{c} & f$_{\rm d}$(C,D)\tnote{b} \\

\multicolumn{1}{c}{region} & \multicolumn{1}{c}{(10$^4$ L$_{\odot}$)} & \multicolumn{1}{c}{(s$^{-1}$)} & \multicolumn{1}{c}{(IRAS)} & \multicolumn{1}{c}{(configuration)} & \multicolumn{1}{c}{(s$^{-1}$)} & \multicolumn{1}{c}{(radio)} & \multicolumn{1}{c}{} & \multicolumn{1}{c}{} \\

\hline
G05.48$-$0.24 & 66.5 & 49.62 & O5   & D & 49.36 & O5.5 & 0.95$^{\rm (1)}$ & 0.45 \\
G05.97$-$1.17 & 15.5 & 48.82 & O6.5 & D & 48.75 & O6.5 & 0.90$^{\rm (1)}$ & 0.16 \\ 
G10.30$-$0.15 & 65.7 & 49.62 & O5   & D & 49.30 & O5.5 & 0.96$^{\rm (1)}$ & 0.52 \\
G12.21$-$0.10 & 86.8 & 49.93 & O4   & D & 49.36 & O5.5 & 0.94$^{\rm (1)}$ & 0.73 \\    
G18.15$-$0.28 & 21.6 & 49.08 & O6   & D & 48.85 & O6.5 & 0.99$^{\rm (2)}$ & 0.42 \\
G19.60$-$0.23 & 14.0 & 48.82 & O6.5 & D & 48.70 & O7   & 0.50$^{\rm (2)}$ & 0.24 \\
G23.44$-$0.21 & 73.9 & 49.62 & O5   & D & 48.95 & O6.5 & NA             & 0.24 \\
G23.71$+$0.17 & 34.3 & 49.36 & O5.5 & D & 49.05 & O6   & 0.95$^{\rm (1)}$ & 0.51 \\
G25.69$+$0.03 & 22.1 & 49.08 & O6   & D & 48.75 & O6.5 & NA             & 0.54 \\
G28.20$-$0.05 & 20.5 & 49.08 & O6   & D & 48.58 & O7   & 0.93$^{\rm (2)}$ & 0.69 \\
G31.39$-$0.25 & 38.9 & 49.36 & O5.5 & D & 48.69 & O7   & 0.99$^{\rm (2)}$ & 0.79 \\
              &       &       &      & C & 48.69 & O7   &                & 0.79 \\
G35.20$-$1.74 & 26.9 & 49.08 & O6   & D & 49.07 & O6   & 0.79$^{\rm (2)}$ & 0.03 \\   
              &       &       &      & C & 49.05 & O6   &                & 0.06 \\
G37.55$-$0.11 & 28.1 & 49.08 & O6   & D & 48.94 & O6.5 & 0.90$^{\rm (1)}$ & 0.13 \\     
G37.87$+$0.40 & 65.1 & 49.62 & O5   & D & 48.98 & O6.5 & 0.44$^{\rm (2)}$ & 0.19 \\
	          &       &       &      & C & 48.96 & O6.5 &                & 0.20 \\  
G45.07$+$0.13 & 49.3 & 49.36 & O5.5 & D & 48.40 & O8   & 0.94$^{\rm (2)}$ & 0.89 \\
G45.12$+$0.13 & 62.6 & 49.62 & O5   & D & 48.79 & O6.5 & 0.71$^{\rm (2)}$ & 0.85 \\ 
G45.45$+$0.06 & 49.2 & 49.36 & O5.5 & D & 49.19 & O6   & 0.96$^{\rm (2)}$ & 0.33 \\
G54.10$-$0.06 & 14.9 & 48.82 & O6.5 & D & 48.49 & O7.5 & 0.99$^{\rm (1)}$ & 0.53 \\
G60.88$-$0.13 &  4.3 & 48.08 & O9   & C & 47.35 & B0   & 0.97$^{\rm (2)}$ & 0.81 \\
G77.96$-$0.01 & 10.2 & 48.62 & O7   & D & 48.24 & O8.5 & 0.96$^{\rm (2)}$ & 0.58 \\
	          &       &       &      & C & 48.04 & O9   &                & 0.74 \\
G111.28$-$0.66&  2.7 & 47.36 & B0   & D & 47.25 & B0   & 0.87$^{\rm (2)}$ & 0.22 \\
	          &       &       &      & C & 47.17 & B0   &                & 0.35 \\

\hline
\end{tabular}

\begin{tablenotes}
\item[a] FIR data were recalculated using the IRAS flux and distances shown in Table~\ref{tab:tab1}. Also a single ionizing star and \citet{Pa73} model were considered.

\item[b] This work.

\item[c] Taken from: (1) \citet{WC89} or (2) \citet{K94}.

\end{tablenotes}
\end{threeparttable}
\end{table*}

For 28 sources, the presence of EE reduces the values of f$_d$, but at different levels (see Table~\ref{tab:tab6}). According to the f$_d$ (configuration D) value reduction, we define these levels as : 1.- \textit{Low}; f$_d$ (configuration D) $<$ 0.42, 2.- \textit{Intermediate}; 0.42 $<$ f$_d$ (configuration D) $<$ 0.7, and 3.-\textit{High}; f$_d$ (configuration D) $>$ 0.7. The f$_d$ (configuration D) value of 0.42 was chosen because it is the lowest value reported by \citet{WC89} and \citet{K94}, while values from f$_d$ = 0.7 to 0.99 were the most typical reported by them. 

In summary, 10 regions (level 1) have f$_d$ (configuration D) $\sim$~0.21, 10 regions (level 2) $\sim$~0.55, and the other 8 (level 3), $\sim$~0.78. The EE increases the N$'_c$ value, raises the $\xi= N'_c / N^*_c$ value and reduces f$_d$. Hence, an under--estimation in the N$'_c$ determination by \citet{WC89} and \citet{K94} was inferred. Table~\ref{tab:tab7} shows the energetic results for the final catalogue. The presence of Extended Emission in \uchii~ regions could help to explain the IR--excess observed as reported and discussed by \citet{WC89} and \citet{K94}, because the N$'_c$ in these surveys with high resolution VLA observations underestimated the ionizing Lyman photons from the EE. The observed extended emission is part of a bigger structure, so, if we consider the total N$'_c$ of the whole region, f$_d$ must drop significantly. Nevertheless, the overestimation of dust in these regions is not realistic as it is observed in \textit{Spitzer} data \citep{Ch06, PH08, F09a, F09b}. Moreover, the presence of stellar clusters in these regions, indicated by both the UC emission and at large scales by the EE, is in agreement with the \citet{K94} conclusion.



\section{Summary and Conclusions}
\label{summary}

\begin{enumerate}

\item We observed extended emission in a sample of 28 \uchiirs. Nine of them were discarded (no EE) and two new sources were added, giving a final catalogue of 21 \uchiies. This catalogue complements the works of \citet{KK01}, \citet{K99}, and \citet{El05} and is the largest sample so far. The EE seems to be common in UC HII regions and deserves special attention in forthcoming studies and analyses. We show that multi--configuration VLA images are critical to confirm the direct connection between \uchiirs~ and their associated EE. 

\item Following the \citet{WC89} morphological classification, we conclude that cometary is the dominant type in our sample (43\%). However, we note in general that cometary arcs, spherical (21.5\%) and irregular with multi--peaked structure (21.5\%) morphologies are common and match with the \textit{Spitzer} emission. MIPS 24~\micron~ and radio--continuum emissions match very well, and better than the other IRAC bands with the radio continuum.

\item The observed extended emission in all sources appears to be arc--min scale complex bubbles, where gas, dust, and both low--mass and high--mass star(s) coexist, as a result of their star formation process. Our morphological study favours the bubbling universe described by \citet{Ch06,Ch07}; \citet[][ and references therein]{Simp12}. Alternatively, a detailed comparison of molecular, IR and radio emission on each source is needed to compare with the predictions of \citet{KK01,KK03} model. The observed radio continuum peaks in most cases are coincident with luminous IR counterparts or can be other \uchii~ regions (saturated at 24~\micron), while the cometary shape can trace PDR$'$s. 

\item The EE helps to explain the IR--excess observed in these regions, because the N$'_c$ calculated in \citet{WC89} and \citet{K94} under estimated the ionizing Lyman photons rate, no matter if they come from a single or several ionizing sources. The assumptions of a single ionizing star and dust--free nebula are not valid in the studies of energetics of \uchii~ regions.

\end{enumerate}

\section*{Acknowledgements}

We are very thankful for the thoughtful suggestions of the anonymous referee that helped to improve our manuscript. The National Radio Astronomy Observatory is a facility of the National Science Foundation operated under cooperative agreement by Associated Universities, Inc. This work is based in part on observations made with the Spitzer Space Telescope, which is operated by the Jet Propulsion Laboratory, California Institute of Technology under a contract with NASA. E. de la F. acknowledges support from CONACyT (Mexico) grant 124449, CONACyT--SNI exp. 1326, PROMEP/103.5/08/4722, and financial support from Coordinacion General de Cooperaci\'on e Internacionalizaci\'on (CGCI), UdeG, via several announcements.


\label{lastpage}

\begin{thebibliography}{99}

\bibitem[\protect\citeauthoryear{Araya et al.}{2002}]{Ar02} Araya E., Hofner P., Churchwell E., Kurtz S., 2002, \apjs, 138, 63 

\bibitem[\protect\citeauthoryear{Bania et al.}{2010}]{Ban10} Bania T.~M., Anderson L.~D., Balser D.~S., Rood R.~T., 2010, ApJL, 718, L106

\bibitem[\protect\citeauthoryear{Bendo et al.}{2008}]{Bdo08} Bendo G.~J., Draine B.~T., Engelbracht C.~W. et al., 2008, \mnras, 389, 629 

\bibitem[\protect\citeauthoryear{Benjamin et al.}{2003}]{Ben03} Benjamin R.~A., Churchwell E., Babler B.~L. et al., 2003, \pasp, 115, 953 

\bibitem[\protect\citeauthoryear{Briggs}{1995}]{Br95} Briggs D., 1995, PhD thesis, New Mexico Institute for Mining and Technology

\bibitem[\protect\citeauthoryear{Carey et al.}{2009}]{Car09} Carey S.~J., Noriega-Crespo A., Mizuno D.~R. et al., 2009, \pasp, 121, 76 

\bibitem[\protect\citeauthoryear{Casoli et al.}{1986}]{Ca86} Casoli F., Dupraz C., Gerin M., Combes F., Boulanger F., 1986, \aap, 169, 281

\bibitem[\protect\citeauthoryear{Churchwell, Walmsley \& Cesaroni}{1990}]{CWC90} Churchwell E., Walmsley C.~M., Cesaroni R., 1990, \aap, 83, 119

\bibitem[\protect\citeauthoryear{Churchwell}{2002}]{Ch02} Churchwell E., 2002, \araa, 40, 27

\bibitem[\protect\citeauthoryear{Churchwell et al.}{2004}]{Ch04} Churchwell E., Whitney B.~A., Babler B.~L. et al., 2004, \apjs, 154, 322 

\bibitem[\protect\citeauthoryear{Churchwell et al.}{2006}]{Ch06} Churchwell E., Povich M.~S., Allen D. et al., 2006, \apj, 649, 759 

\bibitem[\protect\citeauthoryear{Churchwell et al.}{2007}]{Ch07} Churchwell E., Watson D.~F., Povich M.~S. et al., 2007, \apj, 670, 428

\bibitem[\protect\citeauthoryear{Churchwell et al.}{2009}]{Ch09} Churchwell E., Babler B.~L., Meade M.~R. et al., 2009, \pasp, 121, 213 
\bibitem[\protect\citeauthoryear{Condon et al.}{1998}]{Co98} Condon J.~J., Cotton W.~D., Greisen E.~W., Yin Q.~F., Perley R.~A., Taylor G.~B., Broderick, J~J., 1998, \aj, 115, 1693

\bibitem[\protect\citeauthoryear{Cyganowski et al.}{2008}]{Cy08} Cyganowski C.~J., Whitney B.~A., Holden E. et al., 2008, \aj, 136, 2391-2412

\bibitem[\protect\citeauthoryear{Davidson \& Harwitt}{1967}]{DH67} Davidson K, Harwitt M, 1967, \aj, 148, 443

\bibitem[\protect\citeauthoryear{de la Fuente}{2007}]{F07} de la Fuente E., 2007, PhD thesis, Departamento de F{\'{\i}}sica, CUCEI, Universidad de Guadalajara, M\'exico 

\bibitem[\protect\citeauthoryear{de La Fuente et al.}{2009a}]{F09a} de La Fuente E., Kurtz S.~E., Kumar M.~S.~N., Franco, J., Porras A., Kemp S.~N., Franco-Balderas A., 2009a, Astrophysics and Space Science Proceedings, 7, 167

\bibitem[\protect\citeauthoryear{de La Fuente et al.}{2009b}]{F09b} de La Fuente E., Porras A., Grave J.~M.~C. et al., 2009b, \rmxaa~Conf. Series, 37, 13

\bibitem[\protect\citeauthoryear{de la Fuente et al.}{2018}]{F18} de la Fuente E., Trinidad M.~A., Porras A., Rodr{\'{\i}}guez--Rico C., Araya E.~D., Kurtz S., Hofner P., Nigoche--Netro A., 2018, \rmxaa, 54, 129 

\bibitem[\protect\citeauthoryear{D{\'{i}}az--Miller, Franco, \& Shore}{1998}]{DMFSh98} D\'{i}az--Miller R.~I., Franco J., Shore S.~N., 1998, \apj, 501, 192

\bibitem[\protect\citeauthoryear{Ellingsen et al.}{2005}]{El05} Ellingsen S.~P., Shabala S.~S., Kurtz S.~E., 2005, \mnras, 357, 1003

\bibitem[\protect\citeauthoryear{Fazio et al.}{2004}]{Faz04} Fazio G.~G., Hora J.~L., Allen L.~E. et al., 2004, \apjs, 154, 10

\bibitem[\protect\citeauthoryear{Flagey et al.}{2006}]{Fla06} Flagey N., Boulanger F., Verstraete L., Miville Desch{\^e}nes M.~A., Noriega Crespo A., Reach W.~T., 2006, \aap, 453, 969

\bibitem[\protect\citeauthoryear{Flagey et al.}{2011}]{Fla11} Flagey N., Boulanger F., Noriega-Crespo A., Paladini R., Montmerle T., Carey S.~J., Gagn\'{e} M., Shenoy S., 2011, \aap, 531, A51 

\bibitem[\protect\citeauthoryear{Franco et al.}{2000}]{Fr00} Franco J., Kurtz S., Hofner P., Testi L., Garc\'{i}a--Segura G., Martos M., 2000, ApJL, 542, L143

\bibitem[\protect\citeauthoryear{Franco et al.}{1990}]{Fr90} Franco J., Tenorio--Tagle G., Bodenheimer P., 1990, \apj, 349, 126.

\bibitem[\protect\citeauthoryear{Garay et al.}{1998}]{Ga98} Garay G., Moran J.~M., Rodr{\'{\i}}guez L.~F., Reid M.~J., 1998, \apj, 492, 635

\bibitem[\protect\citeauthoryear{Gregory et al.}{1996}]{Gr96} Gregory P.~C., Scott W.~K., Douglas K., Condon J.~J., 1996, \apjs, 103, 427 

\bibitem[\protect\citeauthoryear{Hoare et al.}{2007}]{Ho07} Hoare M.~G., Kurtz S.~E., Lizano S., Keto E., Hofner P., 2007, Protostars and Planets V, 181 

\bibitem[\protect\citeauthoryear{Hollenbach \& Tielens}{1997}]{HT97} Hollenbach D.~J., Tielens A.~G.~G.~M, 1997, \araa, 35, 179

\bibitem[\protect\citeauthoryear{Israel, Habing \& de Jong}{1973}]{IHJ73} Israel F.~P., Habing H.~J., de Jong T., 1973, \aap, 27, 143

\bibitem[\protect\citeauthoryear{Kim \& Koo}{2001}]{KK01} Kim K.-T., Koo B.-C., 2001, \apj, 549, 979

 \bibitem[\protect\citeauthoryear{Kim \& Koo}{2002}]{KK02} Kim K.-T., Koo B.-C., 2002, \apj, 575, 327

 \bibitem[\protect\citeauthoryear{Kim \& Koo}{2003}]{KK03} Kim K.-T., Koo B.-C., 2003, \apj, 596, 362

\bibitem[\protect\citeauthoryear{Koo et al.}{1996}]{Ko96} Koo B.-C., Kim K.-T., Lee H.-G., Yun M.-S., Ho P.~T.~P. 1996, \apj, 456, 662

\bibitem[\protect\citeauthoryear{Kumar \& Grave}{2007}]{Ku07} Kumar M.~S.~M., Grave J., 2007, \aap, 472, 155

\bibitem[\protect\citeauthoryear{Kurtz et al.}{1994}]{K94} Kurtz S., Churchwell E., Wood D.~O.~S., 1994, \apjs, 91, 659 

\bibitem[\protect\citeauthoryear{Kurtz et al.}{1999}]{K99} Kurtz S.~E., Watson A.~M., Hofner P., Otte B., 1999, \apj, 514, 232 

\bibitem[\protect\citeauthoryear{Li \& Draine}{2001}]{LiD01} Li A., Draine B.~T., 2001, \apj, 554, 778

\bibitem[\protect\citeauthoryear{Mezger et al.}{1967}]{Me67} Mezger P.~G., Altenhoff W., Schraml J., Burke B.~F., Reifenstein E.~C.~III, Wilson T.~L., 1967, ApJL, 150, L157

\bibitem[\protect\citeauthoryear{Morales et al.}{2013}]{Mo13} Morales E.~F.~E., Wyrowski F.~S., Schuller F., Menten, K.~M., 2013, \aap, 560, 76

\bibitem[\protect\citeauthoryear{Panagia}{1973}]{Pa73} Panagia N., 1973, \aj, 78, 929 

\bibitem[\protect\citeauthoryear{Peeters et al.}{2005}]{Peet05} Peeters E., Tielens A.~G.~G.~M., Boogert A.~C.~A., Hayward T.~L., Allamandola L.~J., 2005, \apj, 620, 774 

\bibitem[\protect\citeauthoryear{Phillips \& Ramos--Larios}{2008}]{PH08} Phillips J.~P., Ramos-Larios G., 2008, \mnras, 391, 1527

\bibitem[\protect\citeauthoryear{Povich et al.}{2007}]{Pov07} Povich M.~S., Stone J.~M., Churchwell E. et al., 2007, \apj, 660, 346 

\bibitem[\protect\citeauthoryear{Rela{\~n}o \& Kennicutt}{2009}]{RelKen09} Rela{\~n}o M., Kennicutt R.~C.~Jr., 2009, \apj, 699, 1125 

\bibitem[\protect\citeauthoryear{Rieke et al.}{2004}]{Rie04} Rieke G.~H., Young E.~T., Engelbracht C.~W. et al., 2004, \apjs, 154, 25 

\bibitem[\protect\citeauthoryear{Ryle \& Downes}{1967}]{RD67} Ryle M., Downes D., 1967, ApJL, 148, L17

\bibitem[\protect\citeauthoryear{Sewi{\l}o}{2006}]{Se06} Sewi{\l}o M.~M., 2006, PhD thesis, The University of Wisconsin -- Madison

\bibitem[\protect\citeauthoryear{Sewi{\l}o et al.}{2008}]{Se08} Sewi{\l}o M., Churchwell E., Kurtz S., Goss W.~M., Hofner P., 2008, \apj, 681, 350

\bibitem[\protect\citeauthoryear{Simpson et al.}{2012}]{Simp12} Simpson R.~J., Povich M.~S., Kendrew S. et al., 2012, \mnras, 424, 2442 

\bibitem[\protect\citeauthoryear{Skrutskie et al.}{2006}]{Sk06} Skrutskie M.~F., Cutri R.~M., Stiening R. et al., 2006, \aj, 131, 1163

\bibitem[\protect\citeauthoryear{Smith et al.}{2006}]{Sm06} Smith H.~A., Hora J.~L., Marengo M., Pipher J.~L., 2006, \apj, 645, 1264 

\bibitem[\protect\citeauthoryear{Stecklum et al.}{1998}]{Steck98} Stecklum B., Henning T., Feldt M., Hayward T.~L., Hoare M.~G., Hofner P., Richter S., 1998, \aj, 115, 767

\bibitem[\protect\citeauthoryear{Tielens et al.}{2004}]{Tie04} Tielens A.~G.~G.~M., Peeters E., Bakes E.~L.~O., Spoon H.~W.~W., Hony S., 2004, Star Formation in the Interstellar Medium: In Honor of David Hollenbach, 323, 135 

\bibitem[\protect\citeauthoryear{Urquhart et al.}{2009}]{Ur09} Urquhart J.~S., Hoare M.~G., Purcell C.~R. et al., 2009, \aap, 501, 539 

\bibitem[\protect\citeauthoryear{Wakker \& Schwarz}{1988}]{Wa88} Wakker B.~P., Schwarz U.~J., 1988, \aap, 200, 312

\bibitem[\protect\citeauthoryear{Watson et al.}{1997}]{Wa97} Watson A.~M., Coil A.~L., Shepherd D.~S., Hofner P., Churchwell E., 1997, \apj, 487, 818

\bibitem[\protect\citeauthoryear{Watson et al.}{2008}]{Wat08} Watson C., Povich M.~S., Churchwell E.~B. et al., 2008, \apj, 681, 1341 

\bibitem[\protect\citeauthoryear{Watson, Hanspal, \& Mengistu}{2010}]{Wat10} Watson C., Hanspal U., Mengistu A., 2010, \apj, 716, 1478 

\bibitem[\protect\citeauthoryear{Wood \& Churchwell}{1989}]{WC89} Wood D.~O.~S., Churchwell E., 1989, \apjs, 69, 831

\end{thebibliography}
\end{document}